# An International Consortium for Evaluations of Societal-Scale Risks from Advanced AI

Ross Gruetzemacher[1,2,3,*], Alan Chan[4], Kevin Frazier[5,6], Christy Manning[1,2], Štěpán Los[2,7], James Fox[8], José Hernández-Orallo[3,9,10], John Burden[3,10], Matija Franklin[11], Clíodhna Ní Ghuidhir[12], Mark Bailey[13,15], Toby Pilditch[2,8,11], Daniel Eth[14], Siméon Campos[15], Kyle Kilian[2,16]

1 Wichita State University, 2 Transformative Futures Institute, 3 Centre for the Study of Existential Risk, University of Cambridge, 4 Mila, 5 St. Thomas University, Miami, 6 Legal Priorities Project, 7 University of St Andrews, 8 University of Oxford, 9 Universitat Politècnica de València, 10 Leverhulme Centre for the Future of Intelligence, University of Cambridge, 11 University College London, 12 Apollo Research, 13 National Intelligence University, U.S.A., 14 Independent Consultant, 15 SaferAI, 16 Center for the Future Mind, Florida Atlantic University

## Abstract

Given rapid progress toward advanced AI and risks from frontier AI systems—advanced AI systems pushing the boundaries of the AI capabilities frontier—the creation and implementation of AI governance and regulatory schemes deserves prioritization and substantial investment. However, the status quo is untenable and, frankly, dangerous. A regulatory gap has permitted AI labs to conduct research, development, and deployment activities with minimal oversight. In response, frontier AI system evaluations have been proposed as a way of assessing risks from the development and deployment of frontier AI systems. Yet, the budding AI risk evaluation ecosystem faces significant coordination challenges, such as limited diversity and independence of evaluators, suboptimal allocation of effort, and perverse incentives. This paper proposes a solution in the form of an international consortium for AI risk evaluations, comprising both AI developers and third-party AI risk evaluators. Such a consortium could play a critical role in international efforts to mitigate societal-scale risks from advanced AI, including in managing responsible scaling policies and coordinated evaluation-based risk response. In this paper, we discuss the current evaluation ecosystem and its shortcomings, propose an international consortium for advanced AI risk evaluations, discuss issues regarding its implementation, discuss lessons that can be learned from previous international institutions and existing proposals for international AI governance institutions, and finally, we recommend concrete steps to advance the establishment of the proposed consortium: solicit feedback from stakeholders, conduct additional research, conduct a workshop(s) for stakeholders, create a final proposal and solicit funding, and create a consortium.

* Corresponding author: ross@transformative.org

# Reading Guide

This is a longer document, and different approaches to reading might be more efficient depending on your background and interest:

- **Full read:** There is a lot of valuable content, including in the footnotes, so, particularly interested readers, and those for whom this work is especially relevant, are encouraged to read the paper in its entirety. For those who do not fall into these categories, the alternative approaches listed below may be useful.

- **Two-minute read:** Review the Abstract and see Figure 1 (page 5), depicting the coordination problem given the existing evaluations ecosystem.

- **Ten-minute read:** Review the Abstract, Figure 1 (page 5), depicting the coordination problem given the existing evaluations ecosystem, skim the list of objectives in S3.1, skim S3.2.1 and review Figure 2 (page 22), skim S4.4, skim S4.5, and review the final section, S6, describing the plan for action.

- **Evaluators and AI risk researchers:** Review the Abstract, skim S2.2, paying special attention to S2.2.5, skim S3.1, review S3.2.1 through S3.2.4, especially Figure 2, skim S4.4, skim S5, especially the second paragraph, and review S6 describing the plan of action.

- **Policy or national security professionals:** Review the Abstract, skim S2.1 and S2.2, paying special attention to S2.2.2 through S2.2.5 on problems with the current ecosystem, skim S3.1 and S3.3, skim S4.1 to S4.5, with emphasis on S4.4, review S5, and skim S5.1 and S6, describing the plan of action.

# Table of Contents

# 1 Introduction

Existing AI and next-generation frontier AI systems[1] pose many serious societal-scale risks (Critch and Russell 2023; Hendrycks et al. 2023; Gutierrez et al., 2023; Bubeck et al. 2023). In light of these risks, scholars and industry leaders have discussed visions for AI governance and regulatory regimes (Anderljung et al, 2023; Ho et al. 2023; Trager et al. 2023; Suleyman and Bhaskar 2023) and have identified specific functions necessary to evaluate and respond to extreme risks from frontier AI systems (Shevlane et al. 2023; Anthropic 2023).[2]

One such function—thorough and continuous risk evaluations[3] of frontier AI systems—deserves significant attention from all stakeholders interested in the effective regulation of these systems. Current approaches to building general-purpose AI systems have tended to produce increasingly large models with surprising and unforeseen capabilities (Ganguli et al., 2022; Wei et al. 2022; Bommasani et al. 2021), and it is anticipated that future emergent capabilities will pose severe societal-scale risks to humans (Bengio 2023; Davidson 2023; Anthropic 2023a). A 2023 survey of governance experts revealed that, out of fifty potential policy interventions for reducing risks from advanced AI, the two most highly prioritized options were pre-deployment risk assessment and dangerous capabilities evaluations (Schuett et al. 2023).

The key role of risk assessment and evaluation in identifying and mitigating advanced AI risk merits prioritization in regulatory efforts. Efforts to assess and evaluate the risks of frontier AI systems can contribute to mitigating societal-scale risk from advanced AI (Shevlane et al. 2023), facilitating effective AI governance (Anderljung et al. 2023), and developing a regulatory framework for approving model training and release (Avin 2023). Regulatory frameworks for such risk evaluations are crucially time-sensitive as capability improvements continue at an accelerated pace, quickly approaching critical thresholds that could result in deaths of thousands or damages in the hundreds of billions of dollars

[1] We define *frontier AI systems* as the most capable AI systems, such as the foundation models (Bommasani et al. 2021) that are pushing the boundaries of state-of-the-art capabilities, including potentially dangerous capabilities sufficient to pose significant societal-scale risks.

[2] The importance of testing and evaluation of AI systems is becoming better understood across a variety of sectors. There are considerable efforts by military organizations to create robust testing and evaluation protocols, and these efforts recognize the crucial role of rigorous and robust risk evaluation of frontier AI (National Academies of Sciences, Engineering, Medicine 2023). Other organizations, like the Center for Strategic and International Studies (CSIS), focus on the importance of evaluation and testing in the development of trustworthy AI systems (Chin 2023; Habuka 2023).

[3] What we discuss here as risk evaluations is equivalent to what others have described as (AI system) audits (Suleyman and Bhaskar 2023). However, AI system audits are only one form of auditing necessary for AI developers—particularly frontier AI developers. Governance audits and application audits are also important for a holistic safety auditing approach (Mökander et al. 2023).

(Anthropic 2023a).[4] This paper proposes the establishment of an organization tasked with coordinating risk evaluations of frontier AI systems and, in turn, producing standards intended to mitigate emergent or unanticipated risks from those systems.

A consortium of frontier AI risk evaluators is needed to address status quo coordination problems[5] between third-party evaluators, AI labs, regulators, and other stakeholders.[6] We illustrate this coordination problem in Figure 1, and provide a more detailed discussion of the need for the proposed consortium in the following section.

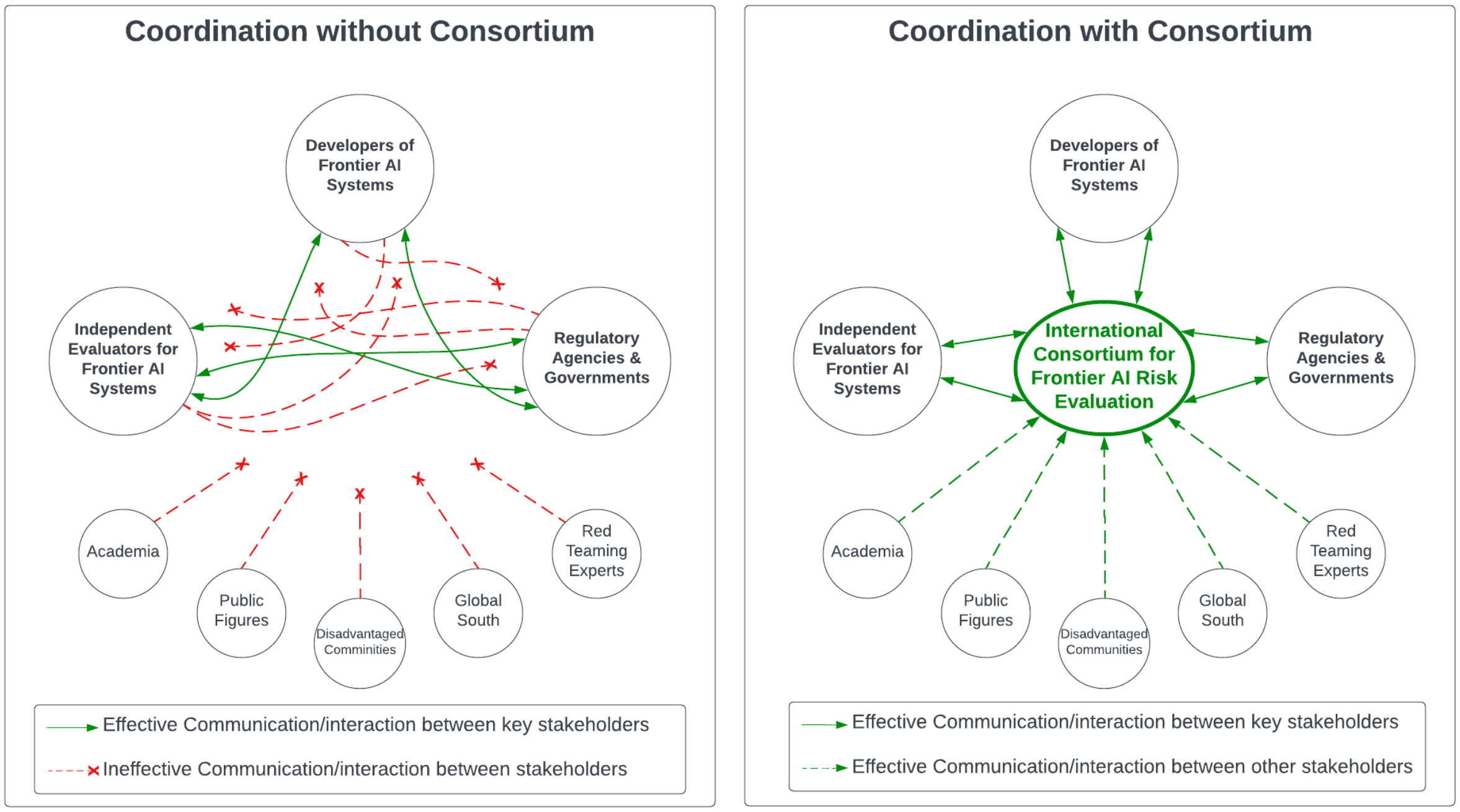

**Figure 1: (Left)** This illustrates the default complexity of communication between large numbers of stakeholders that would result without the creation of an intermediary organization. Many critical elements of the frontier AI regulatory framework may struggle to function effectively without sufficient coordination between the three primary groups of stakeholders. **(Right)** The complex coordination amongst the stakeholders of the frontier AI regulatory framework is simplified and made tractable with the introduction of an international consortium for frontier AI risk evaluation.

[4] In a U.S. Senate committee hearing, Anthropic CEO Dario Amodei warned that AI will likely precipitate grave terrorist threats from biological and chemical weapons between 2025 and 2026 (U.S. Senate 2023); see Boiko et al. (2023) for scientific context. RAND has also described such risks from biological weapons (Mouton et al. 2023). Alternatively, autonomous replication and adaptive systems could lead to highly dangerous cyber threats that are difficult or potentially unreasonable to manage once deployed (Kinniment et al. 2023).

[5] International governance efforts for AI are fragmented, and there exists a broader coordination problem (Gutierrez 2023).

[6] There are analogous institutions in other sectors that bring together the regulators and other multinational stakeholders in a fashion similar to that proposed. Some examples are discussed briefly in a subsequent section.

The remainder of the paper proceeds by first outlining the challenges posed by the status quo regarding the effective mitigation of societal-scale risks by AI systems evaluators. We then identify functions that could be used to address these challenges and review extant organizations with similar functions that can act as examples to draw upon for structuring and organizing the consortium. Finally, we outline the next steps for realizing the consortium.

# 2 The Need for a Consortium

We first provide background on the current state of the frontier AI risk evaluation ecosystem and outline the key problems with the status quo.

## 2.1 The Current Ecosystem

Previous work on the evaluation of risks from AI systems has focused on a variety of risks, ranging from existing harms from AI (Liang et al. 2023; Weidinger et al. 2022) to risks associated with dangerous capabilities of frontier AI systems (Shevlane et al. 2023). Additionally, there are structural risks involving the interactions of AI systems with powerful civil, political, or economic forces in society (Zwetsloot and Dafoe 2019), yet little work exists regarding the evaluation of such risks aside from economics and education. Here, we focus on the evaluation of **societal-scale AI risks**, which we define as AI-precipitated risks to large-scale social systems (e.g., global supply chains, financial systems, or geopolitical stability) or to nations or other large groups if the negative outcomes (e.g., hot war, human rights violations, or economic harms) from these risks are sufficiently widespread.[7,8,9,10,11] Consequently, societal-scale AI risks include far-reaching existing harms, extreme AI risks[12], and structural risks[13]; however, third-party risk evaluators working with the frontier AI labs developing the most capable AI and highest-risk AI systems appear to be focused exclusively on evaluating extreme AI risks and neglect other societal-scale AI risks. National governments have come forward with risk management frameworks for "mapping, measuring, and managing" risks from AI (Tabassi 2023; OECD, 2022).[14] However, the frameworks are designed only as roadmaps for organizations to develop internal evaluation procedures, and do not account specifically for risks from frontier AI.

[7] This definition is derived largely from Critch and Russell's definition of societal-scale harms (Critch & Russell 2023).
[8] While many AI risks constitute societal-scale risks, not all risks fall into this category. Consider risks from autonomous vehicles—such risks would not generally constitute societal-scale risks because the negative outcomes will not be sufficiently widespread. For example, malfunctioning AI-powered autonomous vehicles will, in worst case scenarios, pose risks equal to those of human drivers. Another example could be AI-driven industrial control systems, where normal accidents might be expected, and the scope of the risks would be limited to those working directly with systems involved. If the systems involved are widely used, other users will discontinue their use before the risk becomes sufficiently widespread as to constitute a societal-scale risk.
[9] Evaluation of risks that fall outside our definition of societal-scale AI risks, as well as risks from narrow AI systems constituting societal-scale AI risks, still require evaluation, but, as Mökander et al. (2023) suggest, we feel that this is would be a different layer of evaluation and would need to be considered separately. We later discuss reasons why this might be expected to be the case with AI, and specifically, foundation models, a general purpose technology (Bommasani et al. 2021).
[10] The risks we ascribe to societal-scale AI risks are broader in scope than risks suggested for model auditing by others, including elements like bias and discrimination that have been suggested as appropriate for application audits (Mökander et al. 2023).
[11] This definition would include broader systemic risks such as with the future of work or general cognitive atrophy.

Ethic and safety teams at frontier AI labs appear to be emphasizing a broader range of risks[15] (Anthropic 2023b; Raji et al. 2022b), including viewing AI safety through a sociotechnical lens (Lazar and Nelson 2023), which includes evaluations related to previously neglected structural risks (Weidinger et al. 2023).[16] We use the term societal-scale AI risks in this paper to be inclusive of this broader class of risks, including risks arising from capabilities[17], human interactions, and systematic impacts—the three layer framework of evaluating advanced AI systems set forth by Weidinger et al. (2023)—when the risks meet the other criteria of our definition of societal-salce AI risks.[18]

Evaluation of frontier AI systems for societal-scale risks is critical to addressing risks from advanced AI and broadly capable AI systems (Anderljung et al. 2023). Indeed, a consensus appears to have emerged regarding the need for risk evaluations as essential measures in any comprehensive frontier AI regulatory framework (Engler et al. 2023; Habuka et al. 2023; Suleyman and 2023; Wallin et al. 2023). Several actors, including AI labs developing frontier AI systems, regulatory authorities, and independent third parties, can conduct such frontier AI risk evaluations. Our discussion focuses primarily on third-party organizations devoted to evaluating frontier AI systems for societal-scale risks given the importance placed on such evaluations by frontier AI regulatory frameworks (Avin 2023).

---

[12] Shevlane et al. (2023) define extreme risks as "extremely large in scale…in terms of…impact (e.g. damage in the tens of thousands of lives lost, hundreds of billions of dollars…) or the level of adverse disruption to the social and political order."

[13] Our definition of societal-scale AI risks comprises both short- and long-term AI risks (Sætra and Danaher 2023).

[14] The National Institute of Standards and Technology (NIST) in the United States developed the AI Risk Management Framework (AI RMF), designed for flexibility to address new risks as they emerge since AI impacts are not easily foreseeable. The AI RMF emphasizes the importance of model testing, evaluation, verification, and validation (TEVV). The Laboratoire National de Métrologie et D'essais (LNS) in France is also heavily involved in developing reliable standards and evaluations for AI systems.

[15] While ethics and safety teams appear to emphasize this, they also emphasize that capabilities evaluations should be performed primarily by the labs internally, with minimal oversight by regulators and third-party evaluators (Weidinger et al. 2023). Weidinger et al. (2023) further suggest that application developers should bear the brunt of responsibility for evaluating systems for risks related to human interaction (e.g., misuse), and that third-party stakeholders should bear the responsibility for evaluating systems for systemic risks. This suggestion could be perceived as a not-so-subtle attempt to obviate labs from the consequences of advanced AI systems related to misuse and structural risks.

[16] And a private firm appears positioned to assess capabilities and risks as required by customers, as detailed below.

[17] Weidinger et al. 2023 talk extensively about evaluating capabilities, but do not define capabilities.Similarly, Anthropic (2023) and ARC Evals (2023) do not define capabilities. Shevlane et al. (2023) provide examples of dangerous capabilities, i.e., skills in deception, cyber offense, or weapons design, but they also do not give a definition. Ganguli et al. (2022).do not define capabilities, but suggest that capabilities could extend to entire domains of competency. We discuss capabilities extensively in this work, and, drawing from threads in previous work, we define capabilities as skills or domains of competency associated with learned human behaviors.

[18] In contrast to Weidinger et al. (2023), we feel that third-party evaluations are necessary for all three layers.

The ideal ecosystem of model evaluators would include a diverse set of actors and institutions to ensure rigorous risk assessment across a broad spectrum of societal-scale risks. Yet, at present, this ecosystem remains underdeveloped. Systems with an absence of organic variation may be in such a place because of the properties of the current system (e.g., incumbent advantages, barriers to entry, cronyism, monopolistic behaviors). In such systems, development of variation cannot come without system change (e.g., anti-monopoly policy, financing initiatives, *independent oversight*). In Table 1, we provide a non-exhaustive list of publicly known third-party evaluators of societal-scale risks as of September 2023, along with their affiliation (e.g., industry) and the specific risks they evaluate.

### 2.1.1 AI Evaluation

AI evaluation is an established research area with origins dating to the Turing test (Turing 1950), predating the emergence of AI as a legitimate academic discipline (McCarthy et al. 1956). Claims were first made of a system capable of passing the Turing test in the 1960s with Weizenbaum's (1966) ELIZA, a rule-based natural language processing (NLP) system. Since then, it has been clear that NLP may be sufficient for passing the Turing test and that the Turing test may not be the best evaluation of machine intelligence or whether machines can think (Searle 1980).

**Table I: Third-Party Frontier AI Risk Evaluators**

| Organization | Research Focus | Partners | Additional |
| --- | --- | --- | --- |
| Alignment Research Center (ARC) | self-proliferation, deception, other | Anthropic, OpenAI, Frontier AI Taskforce | theoretical alignment research |
| Apollo Research | deceptive alignment | unknown | mechanistic interpretability and behavioral evaluations |
| Palisade Research | cyber-offense, self-proliferation, other | unknown | advocacy |
| RAND Corporation | biorisks, cyber risks | Frontier AI Taskforce | policy think tank |
| ScaleAI | factuality, reasoning, responsibility | unknown | data validation, dataset creation |
| Stanford Existential Risk Initiative (SERI) | situational awareness, deception | unknown | existential risk research |

Until recently the general approach to evaluating AI systems focused on estimating performance of the system on a data distribution or problem benchmark, mostly disregarding the type of evaluations in the cognitive sciences (Hernandez-Orallo 2017a). This involved specifying evaluation metrics such as sensitivity, specificity, precision, recall, squared error, etc. However, evaluation of AI systems becomes more challenging when trying to assess how well performance on an evaluation metric will generalize to the real world (Manheim 2023), or identifying the capabilities of the system, understood as latent traits explaining and predicting how well the system will behave in new situations (Hernandez-Orallo 2017b, Burnell et al. 2022).

With the rise of machine learning came the rise of benchmark datasets of examples of specific forms of behavior for evaluating capabilities within a specific domain, often intended to be representative samples of the breadth of distributions of real-world tasks. Most such datasets focused on a very narrow domain, but some were more general, and remained relevant for nearly a decade (e.g., the Winograd Schema; Levesque et al. 2012). Such benchmarks, where a model's performance on a dataset is often assigned a single numeric score, have been critical to the evaluation of progress in deep learning (Deng et al. 2009), particularly in domains such as medical diagnostics and for autonomous driving, since these applications are safety-critical (Burnell et al. 2023). Moreover, aggregate benchmarks[19] have also been critical to assessing recent progress in NLP (Gruetzemacher and Paradice 2022), e.g., the General Language Understanding Evaluation (GLUE; Wang et al. 2018) or SuperGLUE, a 'stickier' version of GLUE (Wang et al. 2019). Some aggregate benchmarks, like the massive multitask language understanding benchmark (MMLU; Hendrycks et al. 2021) or BIG-bench (Srivastava et al. 2022) are still used for evaluating performance of current frontier models, although complaints have been voiced about working with some of these datasets (e.g., BIG-bench; Anthropic 2023b, Wang et al. 2023).

More recently, NLP evaluation has had to pivot away from benchmark datasets toward interactive human evaluations (Adiwardana et al. 2020; Roller et al. 2021) due to aggregate benchmarks' limitations (Raji et al. 2021). This can be advantageous in helping to prevent models from overfitting to a measure instead of the task they were designed for[20]; the notion that a measure ceases to be efficacious once it itself becomes the target of optimization is sometimes referred to as Goodhart's law (Goodhart 1971), and can lead to undesirable outcomes such as models that perform poorly on out-of-distribution datasets (Teney et al. 2020). There are several distinct failure modes for this form of overoptimization on the basis of metrics (Manheim and Garrabrant 2019), and is one reason human evaluations are preferable for assessments of societal-scale risks from foundation models. Another reason suggesting the need for human-based evaluations in NLP is the fact that aggregate metrics make it difficult to anticipate how an AI system might perform in a particular situation, and instance-by-instance evaluation reporting is needed to provide more transparent assessments of systems' capabilities (Brunell et al. 2023). Additionally, human-based evaluations in NLP can be useful as the necessity for prompting current models complicates the use of benchmark datasets (Anthropic 2023b; Liang et al. 2023).

[19] An aggregate benchmark is a collection of benchmark datasets that are aggregated to form a single measure. This is opposed to benchmarks involving a single dataset or bank of questions curated for a specific topic (Lin et al. 2022).

[20] In considering the problem of mesa-optimizers (Hubinger et al. 2019), we need to be especially careful to avoid failures of this type dealing with both the base optimizer and the mesa-optimizer.

### 2.1.2 Private AI Testing and Evaluation

AI evaluation has traditionally focused on estimating performance for one or more tasks, and only more recently on identifying capabilities as latent factors (Shevlane et al. 2023; Weidinger et al. 2023). Oddly, the term "testing" has been less common in AI, despite a more natural association with safety than "evaluation". The primary reason for this is that testing is usually applied to verification and validation processes in engineering, with respect to a *specification* and an intended *purpose*, respectively; this is something that many AI developers, especially those of general-purpose AI systems, often ignore. The term "evaluations" is now en vogue among frontier AI developers in reference to capabilities and risk assessments of frontier AI systems, so this is the language we adopt in this document. We continue to discuss testing and evaluation in this section only because ScaleAI uses the terminology, but even they fail to differentiate between the two terms (ScaleAI 2023).[21][22]

Evaluation and testing are usually included in AI audits (Mökander et al. 2023), specifically, technical audits of AI systems; the two other types of AI audits include governance audits and application audits. Significant work has been conducted related to governance and application audits, and there are AI auditing frameworks that exist in these contexts at use in the private sector.[23] However, evaluations in technical audits play

[21] We also discuss testing and evaluation in the following subsection, although regulators do tend to use the terminology appropriately. However, the remainder of this work exclusively utilizes the term evaluations, in the en vogue sense, to refer to capabilities and risk assessments of frontier AI systems.

[22] By evaluations, we mean empirical assessments of models' performance, capabilities, or risks (Shevlane et al. 2023).

[23] Broadly speaking, an AI audit is a process whereby an auditor evaluates an AI system according to a specific set of criteria and recommends further steps to the auditee based on the findings (Costanza-Chock et al. 2022). AI audits may also be classified based on their focus: (i) technical audits analyze technical components including data and source code, (ii) empirical audits measure the effects of a system by examining inputs and outputs and (iii) governance audits assess whether the right governance policies were chosen in developing the system (GOV.UK 2022). A number of existing organizations have attempted to devise comprehensive frameworks for AI auditing. For instance, the Information Systems Audit and Control Association (ISACA) developed a framework to Control Objectives for Information and Related Technologies (COBIT) which spans a system's entire lifecycle, from data gathering and training to deployment (ISACA 2018). However, auditing of applications and system training is not widespread. Consider, the Institute of Internal Auditors' (IIA) Artificial Intelligence Auditing Framework (IIA, 2018) focuses on helping companies (i) assess their AI strategy, (ii) choose appropriate AI governance policies

an outsized role in any framework of auditing for safety of advanced AI or general-purpose AI systems, and play a key role in mitigating societal-scale risks from AI. Application audits are also relevant to mitigating societal-scale AI risks, but they are not the focus of this paper. In this section we focus the nascent private sector efforts related to testing and evaluation of advanced AI systems.

While AI testing and evaluation can clearly be provided as a private service (Mökander et al. 2023; Clark and Hadfield 2019), the ecosystem for private evaluations is less developed than that for independent nonprofit organizations. ScaleAI is now marketing a platform for foundation model testing and evaluation[24] (ScaleAI 2023), and they have support of U.S. authorities (The White House 2023) and have laid out a very different 'Testing and Evaluations Vision" for their platform than the AI risk evaluations efforts discussed previously.

ScaleAI (2023) has proposed five axes of AI systems' abilities that focus on foundation models—instruction following, creativity, responsibility, reasoning, and factuality—the quality of which their platform is designed to assess. In the context of the axes of systems' abilities, they assess for 'helpfulness' or 'harmlessness'. This assessment process is broken down into three complementary parts: (i) model evaluation, (ii) model monitoring and (iii) red teaming.[25] Model evaluation involves an assortment of checks periodically conducted by both humans and AI systems, assessing the models' capabilities and overall helpfulness. Model monitoring consists in automatized post-deployment monitoring of the system in question, checking primarily for anomalous or problematic responses. Finally, red teaming is carried via iteratively targeting specific harms and weaknesses of the system, eliciting undesirable behaviors in order to update and strengthen the evaluation test battery and patch the system through additional tuning. All of these processes are expected to be carried out continually, with increased frequency ahead of releases of frontier AI systems. In May 2023, the U.S. Government selected ScaleAI's platform for the evaluation of frontier models as part of a private sector testing and evaluation initiative (The White House, 2023).

---

and (iii) eliminate the risk of human error when working with AI; testing and evaluation processes are not discussed, which, perhaps, could be expected as the IIA is not a technical organization.

The broader AI auditing landscape seems to suffer from a number of limitations. There is a lack of a unified rigorous framework for conducting AI audits (Costanza-Chock et al., 2022); there is an outsized focus on internal audits (Raji et al., 2022a), running the risk of lower objectivity and less rigorous auditing standards (Brundage et al., 2020); and the frameworks referenced are heavily focused on AI systems' compliance with company policies or on assessing their delivery of economic value to companies. Thus, in lacking comprehensive risk assessment procedures, it is unclear how these frameworks could translate into the context of frontier systems' risk evaluations, especially concerning potential societal-scale risks.

[24] ScaleAI seems to be the only major firm currently offering a foundation model-focused product in the AI testing and evaluations space (ScaleAI 2023).

[25] There is some overlap with the National Institute for Standards and Technology's (NIST) AI Risk Management Framework (AI RMF; National Institute of Standards and Technology 2023), which may not be surprising given ScaleAI's apparent intention of working with regulators.

What ScaleAI (2023) has proposed is more consistent with earlier notions of AI evaluations and involves minimal focus on the risk. Building on their testing and evaluation framework, they propose a vision for the testing and evaluation ecosystem identifying four groups of institutional stakeholders: frontier AI developers, government and regulators, enterprises and organizations seeking to deploy advanced AI systems, and third-party organizations that service the other three groups of stakeholders (e.g., ScaleAI, organizations listed in Table I).[26] The proposed vision goes further to express that the role of government(s) is (i) to establish guidelines and regulations regarding the training and deployment of frontier AI systems and (ii) to set and enforce the adoption of standards on the use of frontier models within government—this includes significant discussion specifically related to the U.S. government. Clearly, ScaleAI's vision for the testing and evaluation ecosystem is supportive of the U.S. government's risk evaluation efforts, but it does not make an effort to suggest how risk evaluation should be conducted. Yet, despite the apparent limitations, ScaleAI's platform likely can play a constructive role in an evaluation ecosystem explicitly engineered to mitigate societal-scale AI risks.

### 2.1.3 Regulatory Testing and Evaluation

There have been modest efforts by public sector institutions to support responsible AI development through risk management frameworks and support for testing and evaluation. For example, the U.S. National Institute for Standards and Technology's (NIST) AI Risk Management Framework (AI RMF; National Institute of Standards and Technology 2023) provides a comprehensive road map for the identification, measurement, management, and governance of AI risk, covering a broad spectrum of risks to individuals, organizations, and ecosystems (Tabassi 2023). The AI RMF proposes a framework for testing, evaluation, verification, and validation (TEVV) throughout the AI lifecycle. The TEVV process aims to be iterative and adaptive to identify emergent risks as models increase in capability. The AI RMF expands the scope of the OECD's AI Classification Framework released a year prior (OECD, 2022). The OECD's framework is a useful adjunct to the AI RMF, classifying AI risk by system type, dimensions, and opportunity for policy intervention (OECD, 2022). Similarly, the U.S. Department of Energy's AI Risk Management Playbook acts as a reference guide to map AI risk categories to mitigation strategies (U.S. Department of Energy, 2023). With the possible exception of the AI RMF, the various public-sector frameworks fail to comprehensively address societal-scale risks from AI and focus predominantly on near-term dangers of limited scope.

[26] This is not entirely dissimilar to the three groups of stakeholders proposed by Google Deepmind: AI model developers, AI application developers, and third-party stakeholders (Weidinger et al. 2023). In this case, third-party stakeholders would comprise both regulators and third-party evaluators, and seems to minimize their need for involvement in the evaluations process. What is similar is that both ScaleAI and Google Deepmind envision some burden of responsibility for model risks being ascribed to application developers. In practice this could mean that culpability for harms arising from misuse of or accidents from applications developed based on their API would fall to application developers.

The EU's Digital Europe Program put forth a more concrete proposal for large-scale Testing and Experimentation Facilities (TEFs)—specialized virtual and physical environments for developers to test systems in real-world environments (European Commission, 2023). The TEF paradigm is designed to aid in the implementation of the EU's AI Act through regulatory sandboxing and act as a filter between technology and society.[27] Programs such as the TEF, with resource assistance for developers and companies, and frameworks like the AI RMF provide a solid foundation for comprehensive AI testing and evaluation. However, without broad coordination and a high-level overview of the ecosystem, significant challenges remain.

## 2.2 Problems with the Current Evaluations Ecosystem

### 2.2.1 Limited Diversity and Independence of Evaluators

Of the few established third-party evaluators in Table I, only one, the RAND Corporation, has a strong record of objective and robust research. The other third-party organizations conducting frontier AI risk evaluations[28] have strong ties to the AI labs creating and developing the systems they are evaluating.[29] These connections may influence evaluators to deprioritize some societal-scale risks, such as those that involve exacerbating economic and political concentrations of power (Chan et al. 2023). They are also likely to invite scrutiny from future regulatory agencies. Additionally, epistemic diversity may also be severely lacking. Ensuring a diversity of evaluation bodies is key to guarding against conflicts of interest and mitigating the potential for evaluation gaming.[30] The benefit of diversity also extends to differential geographies–national and regional–where the cultural context of the region heavily influences national mindsets, core beliefs, concerns about AI risk, evaluation practices, and risk tolerance (Singh et al. 2023).[31] At the same time, a diversity of evaluators without coordination to avoid races to the bottom can mean that companies shop around for evaluators with the most lax standards.

[27] Regulatory sandboxes in AI is where authorities work with companies to test innovative next-generation AI capabilities that may challenge existing regulatory frameworks.
[28] However, the evaluations ecosystem is rapidly evolving, and the Frontier AI Taskforce appears poised to take on a large role, though it is unclear specifically how this role will take form.
[29] To some degree, it may be difficult to avoid strong ties between evaluators and evaluatees given that evaluating novel risks is a difficult technical machine learning problem and those with the necessary skills often know each other.
[30] Evaluation gaming is where the evaluators optimize for demonstrating safety by privileging specific evaluations. This could be intentional or inadvertent and could cause any evaluation metrics to lose their validity (Fist et al., 2023).
[31] A 2023 study, conducted by George Mason University and the Stimson Center, analyzed 213 AI national strategies from 54 countries and found wide variation in the importance of certain strategies and safety measures for managing AI, with a number of clusters or groupings of countries with shared beliefs. Having broad representation from these different clusters would be beneficial for a snapshot of human beliefs and perspectives (Singh et al. 2023).

Nevertheless, it is difficult to get more diverse parties to engage in risk evaluations of frontier AI systems. Although OpenAI has a public researcher access program, the admission process is opaque. Other frontier AI labs like Anthropic and Google DeepMind do not have such researcher access programs[32], and Google Deepmind goes as far as to suggest that capabilities evaluations are best suited to being conducted by advanced AI developers with minimal involvement from third-party stakeholders (Weidinger et al. 2023).[33] This status quo privileges those evaluators who already have personal connections to the frontier labs. Moreover, a regulatory effort for an emerging technology led exclusively by an unrepresentative set of technical experts is likely to neither receive the public's support nor earn and sustain its trust (Stern 2011). AI expertise increasingly is distributed around the world. Risk evaluations conducted by a globally diverse set of actors have better odds of being a common source of information and of spurring cohesive regulatory efforts (Paglia and Parker 2021).[34]

### 2.2.2 Races to the Bottom

Races to the bottom in safety standards are also a significant risk. Suppose that we have set up a licensing scheme wherein AI systems must be evaluated by an accredited organization before deployment. In such a market, risk evaluators who have lower standards get more customers because they are less demanding and provide the same perceived value. Hence, the markets get captured by some of the worst evaluation organizations.

Whether evaluations are conducted via self-policing, or via developer-paid evaluators, the compromised incentives in a race to the bottom produce a negative feedback loop that acts as a growing structural risk to safe AI development. Though negative feedback loops are often seen as stabilizing forces in multi-agent and complex systems (Meadows, 2008), here such stabilization is a continual pressure to minimize the possibility of needed disruption/correction of AI models. Specifically, as the (compromised) evaluator perceives a personal cost to finding risks (slowed model development progress, lost business), the pressure on those actors is to reduce scrutiny, and interpret ambiguity favorably. Even substantive cultural pressure may only produce tokenism in risk discovery, wherein accuracy values are signaled by highlighting discovered risks that are in fact minor/irrelevant to developer progress rates. Critically, as this pressure continues, two pressure loops build. First, investment in scrutinizing processes and methods decreases over time as apparent need lowers with an absence of risk discoveries, lowering evaluative capabilities. Second, internal norms of unproblematic development become more entrenched, making the (social) penalty to potential risk discovery even greater, thus compounding the already compromised incentives. Moreover, as this (currently ongoing) process continues to coagulate it becomes harder to correct.

[32] See Weidinger et al. (2023) Figure 5.1.
[33] Placing the burden of evaluating misuse and structural risks on other parties could be seen as attempting to avoid liability.
[34] Distrust inherent in the East-West dichotomy may make it challenging for Western countries to conduct verifiable evaluations of frontier AI developers in the East.

Race to the bottom dynamics have also been seen in private audits currently being conducted within labs. *The Washington Post* reported that Facebook suppressed reporting harms they knew were present in their models, even making contrary statements to the public (Oremus 2021). In an *MIT Technology Review* article, Google reportedly compelled a researcher to resign over her decision to publish an article on the harms of large language models (Hao 2020). Such high-publicity incidents create a culture that incentivizes employees to compromise on safety if they value their jobs.

Races to the bottom can also occur across national borders, where countries that regulate less and are less risk-averse capture a great share of the value of frontier AI deployment or attract more companies. Moreover, as many societal-scale risks would be global in nature, their costs will largely be external; thus, countries that relax regulation are more likely to internalize a greater share of the benefits from advanced AI systems. Since there is currently no authoritative voice on frontier AI evaluations and no AI risks evaluation standards, countries could point to the least demanding standards from individual organizations when developing their regulations.

### 2.2.3 Suboptimal Allocation of Effort

The status quo may make it difficult to allocate evaluation effort so as to maximally reduce societal-scale AI risks. As the number of model evaluators grows, there is a risk that the lack of effective coordination results in the unnecessary duplication of work. Such duplication is particularly troublesome in light of the resource constraints—e.g., expertise—that will constrain any one evaluation effort. Yet, given that evaluating societal-scale risks from frontier AI systems is still an early science,[35] coordinated independent verification of results will be important.

[35] Given the nascent state of research on evaluating frontier AI systems (Shevalne et al. 2023), it is likely that much can be learnt from reviewing successes and failures of other fields such as psychometrics, comparative and developmental psychology, the extant body of knowledge on the testing and evaluations of AI, or software testing and other areas of engineering and computer security.

At the same time, research foci should be optimally spread amongst the various types of societal-scale AI risks. For example, it would be suboptimal if half of the frontier AI evaluation organizations were working on deception evaluations but none were working on evaluating other salient risks like concentration of power or long-horizon planning. Therefore, an effort needs to be made to coordinate evaluations in order to ensure that at least one organization is working to advance the science of evaluating each salient risk,[36] whilst guaranteeing the capacity required to evaluate all salient risks for all frontier AI systems having the characteristics (e.g., model size) to require such evaluation. Moreover, one must also ensure that organizations are not incentivized to only focus on more tractable risks—e.g. because they are rewarded for making progress faster—as this would lead to the neglect of more speculative risks[37] and increase the possibility of 'unknown unknowns'.

Additionally, efforts by independent AI labs to create risk evaluation networks could possibly be counterproductive, or less helpful than they would otherwise be, to collective safety efforts by focusing interested volunteers on a single organization.[38] Efforts to increase participation in red teaming or evaluation of AI systems needs to be coordinated by an independent party so that efforts are not concentrated on a single frontier AI lab.

### 2.2.4 Barriers to Knowledge Sharing

There is currently no public communication forum for frontier AI risk evaluations, resulting in three significant consequences. First, policymakers have no authoritative, impartial voice to look to about evaluations when crafting legislation. Second, model evaluators lack an easy way to share information about research results, such as methodological problems or the presence of dangerous capabilities in a model, beyond personal channels, academic publications, or haphazard search through masses of preprints.[39] Third, there are no shared definitions (Maas 2023), common evaluation reporting standards[40], or standards of any type in the nascent evaluations ecosystem. Additionally, the lack of transparency regarding models' training data, models' risks, and models' evaluations are also issues of concern.[41]

[36] This may be tricky, and will require efforts to ensure that talented people are incentivized to work on risks that they are both motivated and well-suited to work on.
[37] Rather, independent evaluators should be encouraged to allocate some resources to exploring more speculative risks.
[38] OpenAI's Red Teaming Network is a potential example of this (https://openai.com/blog/red-teaming-network; OpenAI 2023b).
[39] It could be nice to have both private and public channels for this, due to dueling concerns over security and transparency.
[40] Anthropic describes evaluation reporting and the relationship between evaluator and evaluatee as a significant challenge (Anthropic 2023b).
[41] Additionally, there appears to be an effort by Stanford's Center for Research on Foundation Models and Stanford's Institute for Human-Centered AI to emphasize model transparency, including with respect to risks and their mitigation (Bommasani et al. 2023).

Safely sharing and deploying powerful AI systems is yet another concern. The open source approach is one option whereas it is also possible and not uncommon for models to be kept entirely private. Because each of these has downsides, structured access has been proposed as an approach for safely sharing and deploying systems (Shevlane 2022). However, few use this proposed approach, and much work remains to realize its potential. Additionally, another concern that arises with this and other barriers to knowledge sharing is information security.

### 2.2.5 Coordinated Evaluation-based Risk Response (CERR)

Coordination will be critical to the successful development of safe and responsible AI systems (Gutierrez 2023; Askell et al. 2019), and it will be critical to developing a strong AI risk evaluations ecosystem. One particularly underdeveloped area of research is coordinated evaluation-based risk response (CERR). Alaga & Schuett (2023) propose coordinated pausing as a risk response mechanism for mitigating risks from frontier models.[42] Yet, risk response, such as that proposed by Alaga and Schuett (2023) should not be limited to pausing. While ARC Evals has proposed evaluation-based risk response that would involve restrictions when dangerous capabilities are identified (ARC Evals 2023), there has been little effort to identify the mechanisms through which this could be implemented, what specific actions or restrictions should be taken in response to new and unanticipated emergent capabilities, and how or by whom this could be effectively coordinated.

### 2.2.6 Scalability of Evaluations

Given that increased scaling of model training is likely to require increasingly more rigorous evaluations as frontier AI systems' capabilities increase (Anthropic 2023a), and that the latest Nvidia and Google co-processors will make scaling cheaper (Hobbhahn and Besiroglu 2023), regulation will need to be able to keep up with the speed of the technology driving progress and of the businesses seeking to capitalize on it. Consider that GPT-4 (OpenAI 2023a) was trained with 2020 technology (Patel and Wong 2023), and that the current state-of-the-art co-processors only became available on the cloud—and presumably to many other customers—at the end of summer 2023 (Shah 2023). Moreover, within 18 months, some organizations will have the computational resources to train systems up to two orders of magnitude larger[43] than GPT-4 (Harris and Suleyman 2023). Thus, it is likely that more organizations will be capable of training dangerous frontier systems that require more intensive levels of risk evaluation (Anthropic 2023a), which is why acting quickly (Suleyman and Bhaskar 2023) and coordinating a network of evaluators able to scale to an influx in demand for evaluating frontier AI systems is critical.

[42] They propose a scheme requiring that upon discovering a dangerous capability during an internal or external system risk evaluation, frontier developers would agree to collectively pause training, evaluation, or deployment of the system exhibiting the capability as well as other systems which may also exhibit the emergent capability. They further discuss four practical versions of the scheme, all of which require coordinated collective action.

### 2.2.7 Other Problems with the Current Ecosystem

Current evaluation techniques have some inherent problems (Weidinger et al. 2023) that should be considered before scaling to frontier models (Anthropic 2023b). Methods that utilize humans as red-teamers, labelers, or jailbreakers, for example, may be subject to human bias (Weidinger et al. 2021; Huang et al. 2023); methods that rely on machine learning should ensure that the systems used for evaluating other AI systems are trustworthy (Jaiswal et al. 2023); and methods that evaluate for safety can pose safety risks themselves such as leaking data (Ganguli et al. 2022). It could be seen as irresponsible to proceed with scaling evaluations to frontier models without first addressing the lack of reporting standards and standards for risk evaluations necessary to mitigate the dangers of problems of this type. Moreover, these challenges are inherently more difficult to address in light of the lack of a coordinated effort toward establishing the domain of advanced AI risk evaluations as an active and prioritized research discipline.

# 3 An International Consortium: Objectives & Scope

To address the challenges outlined in the previous section, we propose the creation of an **international consortium[44] for advanced AI risk evaluations**. This proposed consortium would coordinate AI risk evaluations amongst the three core groups of stakeholders—developers of frontier AI systems, independent evaluators, and governments and regulatory agencies—providing responsive expert guidance to government regulators in response to the ongoing evolution of risks associated with increasingly large and capable AI systems.

[43] i.e., foundation models that are larger in the number of model parameters or in the amount of compute used during training.

[44] A consortium is a group of—not necessarily homogeneous—entities that come together around a common cause or objective. Consortia can fulfill many different purposes and take a variety of forms, but share a common characteristic: they can accomplish objectives that would be difficult or impossible for the individual members to do on their own (Zavlovinka et al. 2020). This is partially made possible by resource- and knowledge- sharing, risk pooling, enhanced credibility, and establishment of formal relationships (IEEE 2023).

## 3.1 Objectives

We identify three primary objectives for the proposed international consortium:

A. Act as an intermediary between third-party risk evaluators, frontier AI developers, governments and regulatory agencies, and other stakeholders (e.g., academics, civil society groups), including acting to coordinate evaluations-based risk response;
B. To set and implement standards quickly while minimizing bureaucratic challenges;
C. Serve as an advisory body for regulators:
    a. In developing their own risk assessment capacity
    b. In verifying frontier AI evaluators' risk assessment abilities (via certification)

If implemented effectively to achieve these objectives, the proposed consortium would complement some of the previously proposed models (Anderljung et al. 2023; Avin 2023; Suleyman and Bhaskar 2023). However, there are many ways in which a consortium or international institution for model evaluators could be implemented.

On one hand, it is possible that the proposed consortium could come to meet all three of the designated objectives; however, with the UK Frontier AI Taskforce now signaling an intention to take an active role in the evaluations ecosystem, the consortium could alternatively take on a role complementary to the Taskforce or other new organizations (Ho et al. 2023; Trager et al. 2023) to achieve these objectives. In the case of the former, the proposed consortium would provide a simpler and more easily implemented solution for the issues of standard-setting and coordinating evaluations by not requiring that all advanced AI governance functions be housed in a single organization. Yet, in the case of the latter, the proposed consortium could help to reduce the complexity of organizations that have been proposed with a broader mandate including testing and evaluation of risk as a subelement.

## 3.2 Operations

Effectively operationalizing the proposed consortium presents numerous challenges, and we discuss those in this section. We note that a consortium could operate in a fashion that is dissimilar from any existing organization, and we attempt to explain some of these unique operating characteristics that would be beneficial in helping to achieve the objectives outlined above. However, we also note that operationalizing an international consortium like this, for a topic involving a nascent science, is particularly challenging, and so the discussion that follows is far from exhaustive. Further exploration of these topics and engagement with stakeholders will be necessary to realize anything similar to the organization proposed here, and the conclusion of this paper lays out the steps we envision as necessary for doing this.

There are numerous challenges to coordinating evaluation efforts, and there are numerous ways to coordinate evaluations. We focus on some specific issues related to the shortcomings of the existing ecosystem to outline a vision for how a consortium could operate in which it is able to effectively achieve all of the three proposed objectives. While there may be flaws in this framework, we hope that it can spur interest in the virtues of pairing coordination of model evaluations with the standards creation process.

### 3.2.1 Prioritizing Evaluations

Recently, frontier AI labs and risk evaluators have proposed the concept of Responsible Scaling Policies (RSPs; Anthropic 2023a; ARC Evals 2023) that are intended to specify the level of AI capabilities that a frontier AI developer is prepared to safely develop and the threshold at which societal-scale risks would be too great for the developer to continue to deploy systems or to scale systems' capabilities further (Matteucci et al. 2023). This is quite similar to the coordinated evaluation-based risk response (CERR; i.e., coordinated pausing) proposed by Alaga and Schuett (2023) in that both proposals require action when insufficient protocols are in place to proceed with training or deploying a system safely.[45] Moreover, ARC Evals' vision for RSPs sees them as a path toward evaluation-based scaling, which would be a form of CERR.

Anthropic's RSP[46] defines a framework for AI Safety Levels (ASLs) that increase with growing degrees of catastrophic risk[47] (Anthropic 2023a) that will require evaluations capable of detecting warning signs of dangerous capabilities before they reach the next ASL.[48] For example, with our current frontier AI systems we are currently at ASL-2, which means that these systems must be evaluated for ASL-3 warning signs; similarly, ASL-3 systems will need to be evaluated for ASL-4 warning signs, and ASL-$n$ systems will need to be evaluated for ASL-[$n$+1] warning signs. Thus, the framework requires that ASL-[$n$+1] warning signs are defined prior to training ASL-$n$ systems. ASL-4 capabilities and warning sign evaluations have not yet been defined, but Anthropic is committed to defining them before training and evaluating ASL-3 models. This approach attempts to identify emergent capabilities preemptively, which has been suggested to potentially be valuable 'practice' for CERR (ARC Evals 2023).

---

[45] Avin's (2023) blueprint for frontier AI system training could be applied here.
[46] Anthropic's RSP is designed in the spirit of the concept of responsible scaling policies proposed by ARC Evals.
[47] They describe catastrophic risks as the risk of events at the least on the magnitude of thousands of deaths or harms costing hundreds of billions of dollars, similar to the definition of extreme risks by Shevlane et al. (2023). Their RSP is intended to complement their work mitigating other societal-scale risks.
[48] This framework is loosely based on the U.S. government's biosafety levels (U.S. Department Health & Human Services 2015).

While it would be nice if all dangerous emergent capabilities could be determined *a priori*, it is unlikely that this will be the case; it is more likely that some emergent capabilities will be discovered *a posteriori*, necessitating CERR. Therefore, frontier AI systems that are pushing the boundaries of AI systems' capabilities will require more cautious risk evaluations, including substantial red teaming[49], in order to trigger CERR as well as standards creation. The proposed consortium would be able to facilitate these processes while also certifying select AI risk evaluators for this more challenging task of evaluating true frontier AI systems.

Considering this in the context of RSPs and ASLs: models in a lower ASL would pose significantly less risk of emergent capabilities and could be evaluated less cautiously given that the evaluation standards and techniques would have been proven when the previous ASL was at the frontier. However, such models would need to be subject to equally as rigorous evaluations as they would be were they at the frontier, but evaluators would not need to prioritize the detection of dangerous emergent capabilities or red teaming in order to trigger CERR. Advanced AI systems that are not frontier systems but that still require rigorous and robust evaluation may be thought of as **fringe AI systems**.

Determining whether a new advanced AI system would need to be trained as a frontier system or a fringe system could be determined by the proposed consortium, via frequently updated standards and metrics. For current AI systems, the systems' scale[50] would be the primary determining factor for this given its relevance to current frontier AI systems, but other factors could also contribute[51], and both the relative importance of the factors and additional factors could possibly contribute when there are changes in the underlying technologies powering frontier systems. Additionally, scale would determine what constitutes fringe AI systems. At present, frontier AI systems would presumably be all systems equal to or greater in scale than those which fall within Anthropic's (2023a) current ASL-2, but it is still unclear as to what might constitute ASL-3.[52] Using this dichotomy between frontier AI systems and fringe AI systems for prioritizing evaluations, the training of new advanced AI systems could proceed as depicted in Figure 2.

---

[49] Red teaming is critical to successfully navigating risks from frontier AI systems. We note that both Anthropic and OpenAI (and presumably Google and Microsoft) actively engage in red teaming for internal risk evaluations processes (Anthropic 2023c; OpenAI 2023b; OpenAI 2023c).

[50] Assessed as an aggregate measure, capturing dimensions such as the amount of compute (in floating point operations per second hours) used during training.

[51] Maas (2023) describes a lack of consensus regarding the definitions of 'frontier models', and further describes a potential regulatory gap between frontier AI systems and fringe systems, one which we feel that the framework proposed here would ameliorate.

[52] OpenAI's GPT-4 (OpenAI 2023a) and Claude 2 (Anthropic 2023d) would presumably be ASL-2 systems, but it is not apparent where the threshold lies for ASL-3, and whether Google's forthcoming Gemini system (Pichari 2023), a 'next-generation foundation model', would reach this level.

One of the most valuable features of having a consortium manage the evaluations process in a manner such as that described in Figure 2 is that it would allow for quick and nimble standards adoption. For example, it is essential that when dangerous emergent capabilities of a new frontier system are discovered they are quickly shared with other evaluators and frontier AI developers, similar to the sharing of zero-day vulnerabilities among cyber researchers (Suleyman and Bhaskar 2023). Without a consortium, the coordination necessary for CERR interventions, like coordinated pausing (Alaga and Schuett 2023), would be much more challenging and seemingly would still require some sort of nongovernmental organization.

### 3.2.2 Delegation of Evaluations and Accreditation

If frontier AI developers are free to select the organizations that conduct the risk evaluations of the AI systems, this could easily lead to a race to the bottom or to less robust evaluations and red teaming than if multiple organizations were to be involved in the process. To these ends, one possible function for a consortium could be to delegate the responsibilities for evaluating each true frontier AI system.

For one, it is unlikely that a single organization will be able to accomplish a robust, comprehensive, and infallible set of evaluations given the state of the science (Anthropic 2023b). With the present state of uncertainty, diversity and redundancy are the two most powerful tools that can be brought to bear to ensure that the results of risk evaluations for frontier AI systems are reliable and trustworthy. Therefore, it may be beneficial if AI risk evaluations were to be distributed amongst evaluations organizations based on the relative strengths of the organizations. This could be accomplished via a regulatory market approach (Hadfield and Clark 2023) with accreditation in specific types of risk evaluations[53], or, each organization would be required to perform the evaluations delegated to it[54]; after completion of evaluations, each evaluator would be required to produce a report and recommendation regarding its decision to approve or reject the AI system being evaluated. The consortium could provide a layer of anonymity between the evaluators and the frontier AI developers so evaluators would not feel pressured or risk future work in order to disapprove of systems they deem unsafe; however, this may be challenging, and could require evaluators to maintain anonymity regarding their areas of relative expertise, and may not be a realistic option given the close networks of evaluators and evaluatees.

---

[53] The proposed consortium would need to be authorized to certify evaluators in specific types of evaluations. If only a limited number of firms were allowed to be certified for each risk, the competition could prevent a race to the bottom. A race to the bottom may be possible without competition due to the need for red teaming and exploratory risk analysis, and difficulties in accrediting such processes.

[54] This would require price fixing for frontier AI evaluation. Price fixing could be mandated by regulators or frontier AI developers. Alternatively, while U.S. antitrust law applied to non-profit organizations, perhaps exceptions could be made in this circumstance.

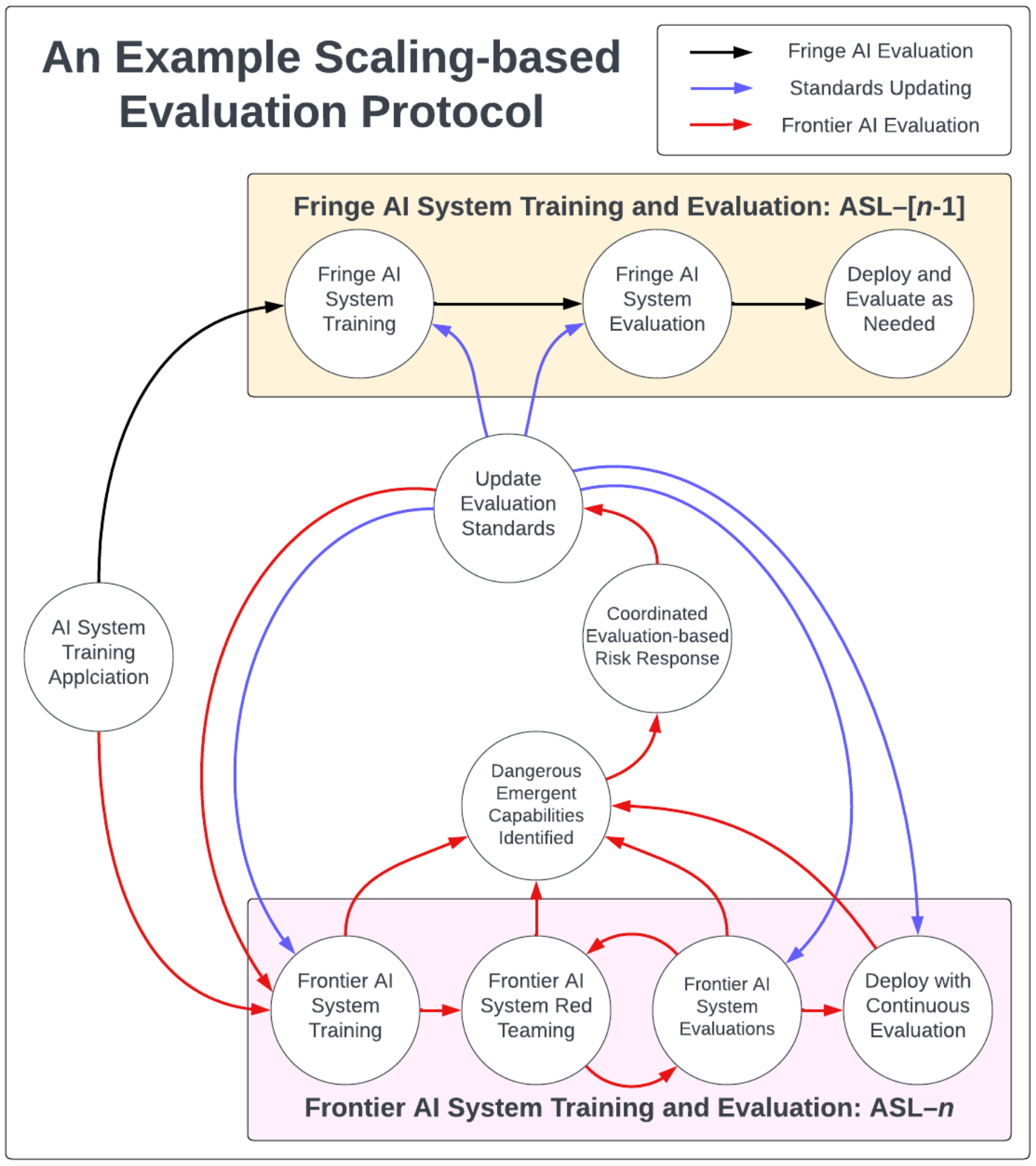

**Figure 2:** AI developers would apply for training with the consortium, and based on whether the proposed system exceeds a certain threshold of computational resources used during training, along with other factors, the system would be determined to be either a frontier system or a fringe system. Frontier systems require more cautious training and evaluation—including a continuous cycle of red teaming and evaluation—based on the anticipated requirements of the next-generation AI Safety Level (ASL; i.e, ASL[$n$+1]) to monitor for dangerous emergent capabilities in order to trigger coordinated evaluation-based risk response and to update evaluation standards; fringe systems are equivalent to systems from a previous ASL (i.e., ASL-[$n$-1]), and, while requiring equally stringent evaluation, do not require evaluators to be cautiously monitoring for dangerous emergent capabilities (e.g., continuous red teaming would not be required during training). When a new AI safety level emerges, all systems of the scale of the previous frontier systems would transition to being trained and evaluated as fringe systems. Note, this figure is an oversimplification of the complex realities of verifiably evaluating fringe systems for societal-scale risks; however, it accurately depicts how frontier AI systems would require prioritization with respect to evaluation rigor, and how information collected from training and evaluating the current ASL as well as all preceding ASLs, could be used to develop quality control protocols that would minimize the oversight required to conduct rigorous and verifiable evaluations of fringe AI systems.

Delegation of evaluation responsibilities, when paired with multiple levels of accreditation for evaluators, would enable allocation of the top talent entirely to frontier AI systems. If further certification was offered for specific types of risk evaluation, and a finite number of certifications were to be available at any time, the competition could prevent race to the bottom dynamics.[55] Private firms, such as ScaleAI, offering platforms for evaluations would likely not be certified for evaluation of frontier AI systems[56], yet, they could still be a very valuable element to the ecosystem in providing a service and platform enabling scaling of evaluations to meet the demand for evaluating fringe systems.[57] Moreover, it is possible that fringe AI systems evaluation could be automated (ScaleAI 2023; Suleyman and Bhaskar 2023).[58]

Challenges could arise if frontier systems requiring evaluation exceeded the supply of certified frontier system evaluators. However, a consortium could work with stakeholders beyond just evaluators, regulators, and AI labs—it could help to coordinate evaluators with academics who could work to increase the supply of talent needed for evaluations in the case of excessive demand. Fortunately, the cost of computational resources needed to scale to the next generation of frontier AI systems is likely to limit the number of actors able to train such systems. If academics were required for evaluations, the consortium could help evaluators quickly identify academics with talent sets that could supplement the talent in their organizations in order to complete thorough and robust evaluations.

### 3.2.3 Evaluator Independence

It will be critical to ensure evaluators' independence with respect to approving systems, determining which evaluators to prioritize, determining how to allocate resources within their organizations, etc. Moreover, evaluators' independence should be thought of as beneficial to risk mitigation efforts in that it fosters specialization among risk evaluation firms. This does present a risk that too many firms choose to specialize in the same type of risk evaluation, but price fixing by the consortium, on frontier system evaluations, would enable the consortium to set prices on all risks such that the space of societal-scale risk evaluations was sufficiently covered by the evaluation organizations certified for frontier risk evaluation.

---

[55] Due to the exploratory nature of frontier AI risk evaluations, such as the need for identifying dangerous emergent capabilities, the need for red teaming, and the need to use ASL-*n* evaluations to set standards for ASL–*n* and to anticipate greater risks for ASL–[*n*+1], competition for specific certifications could enhance rigor and robustness of evaluations. Regardless how robust the accreditation process for evaluators of frontier AI systems might be, it would be challenging to ensure that all evaluators were striving to identify emergent risks via red teaming and exploratory evaluations.

[56] Based on the fact that they do not conduct research to establish new evaluations and that they do not specify evaluation of anything that could obviously constitute a societal-scale AI risk (ScaleAI 2023).

[57] It is not implausible that fringe AI systems will greatly outnumber frontier AI systems given the lower costs associated with training. Moreover, it is unclear whether models fine-tuned on top of frontier models would require treatment as fringe or frontier models. This is a topic that needs further exploration.

[58] This would require trustworthy and verifiably aligned frontier systems, so may be very challenging to implement in practice.

### 3.2.4 Other Considerations

We have described various considerations for the operations of the proposed consortium, but these considerations have focused more on operational matters related directly to risk evaluations. We have ignored other, very important operational concerns.

For one, it is important to consider how the proposed consortium would facilitate membership or go about decision making. Because advanced AI developers and advanced AI risk evaluators are both in positions to conduct high-stakes risk evaluations, and would seemingly be eligible to join the consortium, it is unclear whether it would be beneficial to have both groups be members, at least insomuch as their membership privileges were equal. It could be possible to have members with no conflict of interest have membership privileges in an active sense—e.g., decision making, standards creation—while parties with real or perceived conflicts of interest could also be members, but would have membership privileges in a passive sense.

It would also be very important to identify an effective decision making process for the proposed consortium. One option might be to borrow from the decision making process in use by CERN, which in one capacity involves a high-level executive committee that allows the major players (e.g., funders, states) to make decisions around general research objectives, budgets, etc. Yet, in this model, there is also an internal governance body made up of the actual researchers and experts who exercise substantial control over decisions involving day-to-day operations of the organization.

Additionally, standards creation[59] would be a major function of the proposed consortium. The initial standards creation process would require stakeholders, and likely the inaugural members of the consortium, to agree to standards by consensus. The ongoing standards creation process would be effectively built-in to the evaluation process—if using the process proposed in Figure 2—with CERR triggering standards creation or adaptation when dangerous new emergent capabilities are identified. If regulators are willing to defer to a standards setting body, then this would be compatible with the proposed process and would be a way to avoid a full bureaucratic process and review when integrating new standards.

[59] There need to be standards for both 1) evaluations and 2) evaluations reporting. Standards for evaluations reporting may even be the highest priority given issues Anthropic has expressed having with their attempts to incorporate third-party evaluations (Anthropic 2023b).

Finally, it is unclear how the proposed consortium could be funded. Options include private funding from the firms developing models requiring evaluation; in this case, fees could be charged to frontier AI labs by the third-party evaluators to pay dues or licensing fees to the consortium. Alternatively, governments could subsidize these efforts, but that may be less than optimal as the consortium may be subject to make decisions against its will if under the threat of funding cuts. However, the CERN funding mechanism, discussed below, might be an effective option. Yet another option could be a hybrid, public/private funding model. A discussion of this, as well as other resource-related issues that the proposed consortium would face, are topics that will require further scrutiny and would benefit directly from feedback by stakeholders likely to be involved with the proposed consortium.

## 3.3 Governance

How the proposed consortium will be governed presents numerous challenges also. We discuss these challenges in this section.

The regulatory tasks at issue would benefit from an organizational structure that facilitates the following: *independence, inclusivity, transparency,* and *cooperation*. Though a consortium does not necessarily guarantee these attributes, the flexibility associated with designing such a consortium increases the odds of realizing them. Regardless of whether the consortium forms within a government, in association with governments, or otherwise, these governing values should inform the consortium's operations and structure. While this section briefly discusses why a consortium untethered from any one government may advance the aforementioned values, we do not go as far as to recommend a single "best" approach to organizing such a consortium. In fact, it's likely that no perfect approach exists—the version of the consortium most likely to achieve these governing values will have to adjust its governance mechanisms as the nature of its regulatory tasks change. With that said, the remainder of this subsection presents a high-level overview of why these values merit consideration and how they may manifest in the consortium.

**Independence of the proposed consortium:** whereas an entity situated within an AI Lab or a government may have to respond to the will and whim of actors with motives beyond well-intended risk evaluation research—research with no commercial or militaristic motive—a consortium may benefit greatly from developing a funding mechanism and governance model that ensures researchers can operate in an independent fashion. Which mechanism and model would facilitate that independence, though, remains an open question. For example, CERN's funding mechanism—an annual contribution by member states tied to their gross domestic product (GDP)—is disconnected from the voting power of those member states, with each state still having only a single vote. However, in practice member states with larger GDPs have occasionally attempted to influence CERN operations in proportion to their relative financial contributions to the organization (Sullivan 1987). In the case of CERN, those efforts have largely failed, leaving scientists free to conduct whatever research aligns with the science, rather than the political priorities of the biggest spenders. While such outcomes are not guaranteed, some institutional designs have better odds of protecting the proposed consortium's independence by reducing the influence of biased actors on the caliber and direction of the research inputs and outputs.

**Inclusivity of the proposed consortium:** a consortium unrestrained by national security concerns and by laws limiting participation by non-citizens is theoretically better able to recruit a global set of AI researchers. Consider that only a modest fraction of the world's AI research community would have access to the National AI Research Resource (NAIRR) proposed by Stanford HAI and studied by Congress (NAIRR Task Force 2023). This is a significant limitation that advantages those already with disproportionate research resources. Moreover, it is possible that this lack of inclusivity could reduce the quality of the research and could stifle efforts to increase inclusivity in the broader AI research community. Given the role of the proposed consortium in mitigating societal-scale AI risks, it will be of the utmost importance to foster broad inclusivity. Thus, efforts should be made to ensure that the governance structure of the consortium involves participants from around the world, especially AI experts and government representatives from the Global South, that are often excluded from discussions regarding the global economy or the governance of new emerging technologies.

**Transparency of the proposed consortium:** a consortium again has greater potential to operate in an open and more participatory fashion than other institutional arrangements. Administrative agencies in the U.S., for instance, often struggle to solicit feedback and engagement from a broad array of stakeholders. Comparatively, a consortium could lean into a broader network of actors and employ a set of innovative participatory tools such as deliberative polling and community forums to rapidly solicit feedback and disseminate information in a regulated fashion. Moreover, the consortium would face fewer legal constraints with respect to sharing certain information. However, while maximal transparency would always be the intention, national security or AI safety concerns would likely conflict with this objective. Therefore, it would be important to be transparent about the mechanics of the proposed consortium's transparency, and to have third party audits of the consortium to ensure that the commitment to being as transparent as possible is always actively pursued.

**Fostering international cooperation among diverse stakeholders:** many of the AI governance entities that have been proposed in the U.S. have included an explicit mission to further U.S. aims. Here again, the NAIRR provides a useful case study; Congress authorized a study of the NAIRR (NAIRR Task Force 2023), in part, to "ensure continued U.S. leadership in AI research and development[.]" A consortium oriented, even partially, around the priorities of any specific nation will struggle to foster a cooperative research community—as long as there's a direct connection between a research entity and a nation's ambitions, large swaths of the global research community would have justification to divert their expertise and resources elsewhere.

The aforementioned governing values are easily listed but rarely realized. The few institutions that have managed to embody those values have typically adopted a bespoke funding mechanism and governing model. The consortium, if it embraces these values, will likely need to be structured as an entirely new type of international institution (Maas and Villalobos 2023).

# 4 International AI Institutions

The socioeconomic pervasiveness of advanced AI, along with other fundamental challenges related to AI alignment and international coordination, will make it more challenging to govern than other technologies. In this section we review extant international institutions, proposed AI governance institutions, and extant consortia to glean lessons learned and future guidance and to situate the consortium in the context of other proposed AI models of governance. Based on this survey, we envision and describe two alternative ways in which a consortium or a similar international institution, dedicated to AI risk evaluations, can be realized: through regulatory markets or through the establishment of an intergovernmental organization (e.g., through a treaty, an international agreement, or an act of the United Nations).

## 4.1 Unique Governance Challenges for AI Institutions

One unique governance challenge of advanced AI is that it—specifically, foundation models—is a general purpose technology (GPT; Eloundou et al. 2023; Acemoglu and Johnson 2023; Bommasani et al. 2021). In this way, it is different from many preceding technologies that have been successfully governed, as governing GPTs is fundamentally distinct from prior work proposing new organizations. For example, the internal combustion engine is a widely known GPT (Lipsey et al. 2005). It led to the creation of automobiles, but governing automobiles requires regulating their use, regulating their safety features, regulating their emissions, and enforcing all of these regulations. Additionally, the internal combustion engine also powers over three times more aircraft than jet engines (AOPA 2019), and, to a large degree, this requires separate regulatory agencies than those for automobiles. Yet there are even more domains made possible by the internal combustion engine that also require regulation e.g., motorized watercraft, construction and agricultural equipment. Similarly, electricity spawned numerous domains that each require regulation.

GPTs are fundamentally different to regulate than other technologies because they are characterized by their pervasive impact through all sectors of the economy and their inherent potential for further technical improvements and innovation (Bresnahan and Trajtenberg 1995), and we expect AI to be even more pervasive than past GPTs (Gruetzemacher and Whittlestone 2022). In fact, powerful 'jailbroken' AI systems have already proliferated into the Dark Web (Montalbano 2023). These models lack the controls built into mainstream models like GPT-4, which endeavor to prohibit nefarious use (e.g., malware coding, bomb making) Institutions for regulating traditional industries, such as the Forestry Stewardship Council, regulate the flow of physical goods that only a few companies have access to; similarly, the International Maritime Organization (IMO) and the Federal Aviation Administration (FAA) regulate transportation, which, while not a physical good, nevertheless has scarce physical instantiations that can be export controlled. The pervasiveness of GPTs means that regulation in such a straightforward manner is not possible.[60] For example, anyone with internet access can use GPT-4. Containing and controlling frontier AI suddenly becomes very difficult (Suleyman and Bhaskar 2023), presenting a unique governance challenge. This may require innovative regulatory approaches like regulatory markets (Hadfield and Clark 2023).

Another unique governance challenge of advanced AI is a result of the large number of coordination challenges it poses; as a result, challenges associated with its governance constitute a special class of public policy problems called super wicked problems (Gruetzemacher 2018). Super wicked problems are an extension of wicked problems, which are a class of ill-defined policy problems that have no definitive solution, only satisficing solutions that are only able to be formulated with severe restrictions (Rittel and Webber 1973). Super wicked problems extend this concept to include problems with four additional characteristics: (i) time is running out (ii) there exists no central authority (iii) those seeking to end the problem are also causing it (iv) hyperbolic discounting (Levin et al.

[60] Hadfield and Clark (2023) suggest that because of the nature of AI and its characterization as a GPT, it may be more useful "to think of AI as disrupting regulation itself".

2012). The most common previous example of a super wicked problem is climate change, although it has been suggested that global pandemics may also constitute super wicked problems (Auld et al. 2021). Gruetzemacher (2018) suggests that challenges from advanced AI fit all the descriptions of not one super wicked problem but multiple (e.g., alignment, governance coordination)[61], entangled super wicked problems. The lessons of success, or rather, the lack thereof, on other super wicked problems, like climate change or the COVID19 pandemic, do not bode well for finding a solution to the novel class of entangled super wicked problems posed by advanced AI.

## 4.2 Learning from Existing International Institutions

In this section, we review existing international institutions that offer insights for how the proposed consortium, or some variant thereof, might function. We focus on the implications of the extant organizations on each of the proposed consortium's three objectives. We describe some existing organizations, their activities, and their stakeholders in Table II, and we use the remainder of the section to discuss the relevance of these organizations to the consortium. Yet, there are many limitations to this discussion relevant to organizing and implementing a consortium, for example, how such an organization will be funded or what levers are available to enforce the organization's guidelines and recommendations. A more thorough discussion of the topics touched on in this section is a very important direction for future work.

One core component of any solution to the evaluations coordination problem is the need for coordination between a variety of stakeholders, which, importantly, includes both auditors and auditees. Of the organizations we have reviewed, the International Civil Aviation Organization (ICAO) is the most exemplary because it employs in-house auditors who audit the member countries' aviation capacities, so both the auditors and auditees are members. The Payment Card Industry Security Standards Council (PCI SSC) is very similar because regulatory bodies and standard-setting organizations, as well as the companies that voluntarily adopt these standards, are all members; however, the PCI SSC is concerned with standards and not auditing. The Global Earthquake Model (GEM) is yet another example because it has both policymakers and natural-hazards researchers as members.

[61] The broad challenges of safely aligning AI with human values (Lieke and Sutskever 2023; Christian 2020) and of effectively establishing an international governance over advanced AI (Dafoe 2019; Perry and Uuk 2019) that leads to vibrant and prosperous futures for all of humanity while averting issues from AI race dynamics (Armstrong et al. 2013) or other international concerns can clearly be considered as super wicked problems. However, more narrow forms of advanced AI, such as domain-specific applications of foundation models, could also constitute super wicked problems (e.g., biorisks, cyber risks). The confluence of numerous such problems of this class is what leads to the proposal of a new class of entangled super wicked problems.

## Table II: Existing Organizations with Similar Functions

| Organization | Brief description | Main activities | Main stakeholders |
|---|---|---|---|
| International Civil Aviation Organization (ICAO) | Specialized agency of the United Nations observing the administration & governance of civil aviation | International coordination; establishing guidelines/ standards; conducting compliance audits & evaluations | Signatories of UN Chicago Convention; orgs including civil & professional associations |
| Payment Card Industry Security Standards Council (PCI SSC) | Global forum developing data security standards for safe payments | Conducting compliance audits with PCI standards; standard-setting; professional training; developing new security solutions | Industry companies; national and regional standard-setting organizations |
| Global Earthquake Model (GEM) | Global partnership to reduce risk from earthquakes & natural hazards | Providing data, open-source models, risk assessment software and expertise | Governments; organizations (private, public, professional, nonprofit); & individuals |
| Forum of Incident Response and Security teams (FIRST) | A global forum connecting incident response teams | Standard-setting for cybersecurity; knowledge-sharing between professionals | Private and public incident response teams |
| World Wide Web Consortium (W3C) | Global consortium ensuring the interoperability, safety and transparency of web technologies | Developing and publishing web standards; information exchange between stakeholders | Technology companies; liaison partnerships with civil and professional associations and country governments |
| International Association of Insurance Supervisors (IAIS) | International organization of professionals that covers 97% of global insurance premiums | Insurance standard-setting; works to promote, train, and peer review observance of standards | Insurance firms and partners like the World Bank, the IMF, and the International Organisation of Securities Commissions |
| Intergovernmental Panel on Climate Change (IPCC) | Intergovernmental organization consolidating, verifying, and summarizing climate science research | Organizing large groups of experts to -sort through new climate science research and to produce assessments | UN and WMO member states and global set of researchers |
| The International Police Organization (INTERPOL) | Intergovernmental organization, connecting police teams and law enforcement agencies | Globally connecting and coordinating law enforcement teams; capacity building; running databases and housing expert teams and response teams | Member states' law enforcement teams |
| The Australia Group | An informal forum to ensure states' exports don't aid in the development of chemical or biological weapons | Uphold licensing standards and coordination of national export control measures | Member states (to fulfill obligations of Chemical Weapons Convention and the Biological and Toxin Weapons Convention) |
| The International Atomic Energy Agency (IAEA) | Intergovernmental organization ensuring safe nuclear energy development and monitoring nuclear weapons proliferation | Inspecting nuclear facilities; providing information & developing standards; acting as a hub for knowledge-sharing; ensuring peaceful use of nuclear | Member states of the Nuclear Nonproliferation Treaty |

Another core component of solving the evaluations coordination problem is the flexible development and implementation of standards. Here, the Forum of Incident Response and Security Teams (FIRST) is the most relevant example because it must prioritize the quick adoption of standards in response to new developments in the fast-paced domain of cybersecurity. The World Wide Web Consortium (W3C) is another strong example of an effective standards-setting organization, although it typically operates with less urgency than FIRST. Finally, the International Association of Insurance Supervisors (IAIS) is a strong example as it is a voluntary membership institution of insurance supervisors and regulators that functions as the international standards-setting body for the insurance industry. Additionally, the ICAO and PCI SSC are useful examples of organizations that effectively function to manage standards-setting in addition to fostering multi-stakeholder coordination, achieving the first and second objectives of the proposed organization.

Yet another component of solving the coordination problem requires the certification of frontier AI risk evaluators. The Intergovernmental Panel on Climate Change (IPCC), for instance, does not perform research of its own but collects, verifies, and summarizes the latest climate science research in authoritative assessments. These assessments often serve as the basis for local, subnational, national, and international policy responses to emerging climate risks. Relatedly, with respect to advising government agencies, the International Police Organization (INTERPOL) is an example of an international institution that advises policymakers and national law enforcement bodies in their preparations for future security challenges. The Australia Group, composed of 43 countries, relies on informal agreement among all participants to uphold licensing standards for chemical and biological exports that prevent such goods from being diverted to weapons manufacture. These licensing standards are agreed upon by all members and updated annually. Other international institutions' core functions concern accreditation of auditors. This alone may be a valuable function of a consortium or similar institution in the frontier AI evaluation ecosystem if the effort to create an organization working toward all three proposed objectives fails. Either way, it may be possible to learn from these examples:

- International Audit Practice Consortium (IAPC)
- The International Register of Certified Auditors (IRCA)
- International Organization of Supreme Audit Institutions (INTOSAI)

Another useful example might be the International Atomic Energy Agency (IAEA)—similar to the ICAO as a model of coordination of policy and regulation (Maas and Villalobos 2023)—which is often referenced in the context of frontier model regulation (Ho et al. 2023; Shavit 2023). Additionally, the European Union's establishment of four Testing and Experimentation Facilities (TEFs) for AI (European Commission 2023) could serve as a good model for how academics could be engaged in AI risk evaluations.

These examples—and potentially other examples (Maas and Villalobos 2023)—will be useful to draw on in developing an understanding of the best practices required to meet all three of the objectives designated for the proposed consortium. Future efforts should do this, and build on the cursory analysis described here in order to inform the architects of the proposed consortium when they begin to address the looming challenge of crafting an effective and scalable organizational structure and governance.

## 4.3 Salient Lessons from Previous Proposals

Maas and Villalobos (2023) have reviewed a large number of proposals for international AI institutions covering a wide range of institutional functions. Based on their review, the proposed consortium appears to represent a novel proposal, but one which still aligns with and complements or supplements some of the strongest alternative proposals.

Trager et al. (2023) propose an International AI Organization (IAIO) that takes a jurisdictional approach to AI coordination, whereby jurisdictions such as states or nations must comply with certain safety standards to maintain access to the international AI market. The authors emphasize that a critical prerequisite for this governance structure to work is an international consensus on the regulatory standards. If the capabilities of the models being regulated are not fully understood by all actors, and if agreed-upon mechanisms are not in place to continually monitor capabilities, it is unreasonable to expect to regulate them with any degree of effectiveness. Creating a consensus on standards is beyond the scope of the proposed IAIO, but is one of the core objectives of the proposed consortium. Thus, if a jurisdictional approach were taken, the consortium could own the certification and inspection aspects of jurisdictional compliance determination.

Ho et al. (2023) propose a different framework, expounding on four different institutional models based on their function in governance. The proposed consortium aligns closely with the second institutional model—an Advanced AI Governance Organization (AAIGO). The purpose of the AAIGO is to set regulation standards and promote their enforcement on an international scale. The AAIGO would also be responsible for monitoring compliance and could help member states and participants come into compliance with the standards, perhaps in a way similar to Occupational Safety and Health Administration (OSHA) voluntary inspections. The proposed consortium also aligns to some degree with the first institutional model as well—the Commission on Frontier AI—which would aim to establish a scientific base of knowledge on which the other functions may draw. The consortium would be responsible for establishing the base of knowledge on specifically frontier evaluations. It should be noted that, like Trager et al. (2023), the authors stress the need for consensus as a starting point for any AI initiative.The authors also provide examples of multiple organizations already filling this role: GPAI, the G7 Hiroshima Process on AI, the OECD's AI Principles and AI Policy Observatory, and ITU AI for Good. However, all of these organizations have generally broad foci, and, they are disseminating research on a host of AI-related issues but not specifically evaluating model capabilities. It is possible that given the unprecedented breadth of frontier AI capabilities and the inherent emergent capabilities, robust evaluations and their management could occupy such a formidable amount of bandwidth that organizations with broad foci such as the above would not be equipped to incorporate. This could be another benefit of an entire organization dedicated to evaluations.

In their broad survey of the international AI governance institutions and institutional models that have been proposed to date, Maas and Villalobos (2023) focus on proposals that would involve the creation of completely new international institutions. They identified seven different types of institutional models based on the institutions' general functions. The consortium directly relates to Model 1: scientific consensus-building. An essential role of a scientific consensus-building model is providing a universal, rigorous base of knowledge on which the international community can trust and act, the very purpose of an evaluation consortium. A similar Model 1 organization described is the Global Partnership for AI (GPAI), a collaborative initiative engaging government, industry, and academics. However, GPAI is broad in scope. It disseminates knowledge about a variety of non-technical AI-related topics, promotes policy creation, and invests in AI research. Yet, GPAI is not positioned to coordinate technical evaluations.

The proposed consortium also fills aspects of Model 2—political consensus-building and norm-setting—because it provides the information and credibility to set the norms and aims to harmonize consensus across all of its stakeholder members: industry, government, research, etc. A multitude of AI governance organizations have been proposed in this area, such as the IAIO, and some already exist such as the G7 Hiroshima Process on AI. None specifically identify as consortiums, even though a consortium approach could represent an ideal way to generate consensus across political domains.

The proposed consortium would directly embody Model 3—coordination of policy and regulation—and, Model 4—enforcement of standards or restrictions. Several of the proposed organizations from Models 3 and 4, while they do not explicitly involve evaluations, essentially call upon an evaluations process as an integral part of the governance mechanism; these examples include the International Agency for Artificial Intelligence, the Emerging Technologies Treaty, or the International AI Safety Agency. A consortium is a preferential vehicle for facilitating evaluations as opposed to requiring each of these organizations to conduct their own evaluations because it avoids duplication of effort and affords consensus.

The proposed consortium could also contribute to Model 5—stabilization and emergency response—by setting procedures for such scenarios and operating whistle-blower mechanisms. Additionally, it could contribute to Model 6—international joint research—by conducting research on evaluation methods and goals as a necessary part of its general purpose.

Gutierrez (2023) proposes a single multilateral coordination initiative bringing together all actors and jurisdictions around the subject of mitigating AI risk. This far-reaching initiative would be tasked with identifying risk, coordinating responses, and enforcing agreements. It would consist of membership from states and experts and held accountable to an arbitration board, ultimately to be managed by the United Nations. While this is a centralized approach to governance, it does resemble a consortium approach in that it brings together member states, draws on expert knowledge, and seeks to solve the coordination problem; the difference is that the proposed consortium would focus solely on evaluations and engage a wider variety of stakeholders.

## 4.4 The Benefits of a Consortium

A consortium takes a fundamentally different organizational approach than the existing and proposed AI institutions discussed thus far (Maas et al. 2023; Trager et al. 2023; Ho et al. 2023; Gutierez 2023). Consortia are especially helpful in spaces with many independent actors who need to coordinate action around a shared goal (IEEE 2023), in this case, the goal of robust verification of the safe training and deployment of frontier AI systems.

Certification has been proposed as a key next step in AI governance (Cihon et al. 2021). If the proposed consortium were to successfully begin to certify evaluators, it would effectively create a regulatory market for frontier evaluations whereby it serves as the intermediary. A regulatory market approach was suggested by Hadfield and Clark (2023; Clark and Hadfield 2019) and the approach incorporates the dynamic of regulatory intermediaries (Abbott et al. 2017), whereby instead of regulators directly interacting with the targets of regulation, they act via intermediaries who form the regulatory market. The proposed consortium working in partnership with independent evaluators would together comprise the intermediaries. This would be a unique extension of Abbott et al.'s (2017) regulatory-intermediary-target (R–I–T) model. Rather than the R–I–T framework, this would involve two classes of intermediaries between the regulator and the targets of regulation: R–$I_1$–$I_2$–T. These evaluators would be members of and receive certification from the consortium. Governments could then require models to be evaluated by these evaluators before deployment, even requiring multiple evaluators to evaluate each model and a variety of certifications for the different types of societal-scale risks.[62]

Importantly, certifications could be awarded in a competitive manner, rewarding research and innovative techniques for evaluating different types of risks, effectively fostering the growth of the field of evaluations science. Hadfield and Clark (2023) explain that encouraging investment in novel methods for aligning targets' behavior is a primary goal of the regulatory market approach, one that helps in overcoming the technical deficit of direct government regulation. Competition for a small, finite number of certifications for each of the various types of societal-scale risks being assessed in frontier AI systems could be one means of achieving this goal. Certification for evaluation in each of the risk categories could be determined by diverse panels of experts with no conflicts of interest. This would address a fundamental limitation of the regulatory market approach in that it would mitigate the risk of contracts for risk evaluation being awarded to risk evaluators with close ties to the frontier AI labs being evaluated.

Molding the consortium into a regulatory market approach (Hadfield and Clark 2023) addresses several of the problems with the current evaluation landscape. A field of private evaluators confers the advantage of employing specialized technical knowledge about AI systems that governments lack, conferring the crucial ability to respond more quickly to new developments since they are intimately plugged into the field. Further, a private arrangement fosters competition and innovation, two American economic values which the U.S. has made clear their AI regulatory landscape must support. The arrangement incentivizes evaluators to offer the highest quality, lowest cost, and most cutting-edge services, rather than incentivizing them to cut corners on safety, as they are all held to a high legal safety standard (Farmer et al. 2019). Transparency[63] is key to this holding true, but if implemented properly, this could go a long way toward avoiding race to the bottom dynamics. The consortium then serves as the intermediary between governments and evaluators. Indeed it is hard to envision a regulatory market approach, which relies on intermediaries between regulators and targets being able to effectively provide CERR without a consortium or similar organization coordinating that intermediary space.

[62] Initially there may be a few types of certifications—e.g., for bio risks (Mouton et al. 2023), cyber risks (e.g., Kinniment et al. 2023), and ethical concerns—which could later grow to accommodate certification for evaluating dangerous emergent capabilities or additional risks requiring specialized skills to evaluate.

[63] Specifically, transparency of outcomes to goal-setters in the system and actor awareness of this transparency—the proposed consortium would be poised well to facilitate this.

Consortia have been successful performing similar roles to regulatory market intermediators in other fields. The Geosynthetic Institute (GSI) is an example of a consortium in the field of geotextiles, which are fabrics used in soil applications such as stabilization, drainage, or filtering. GSI membership consists of private companies in a variety of fields such as energy and waste management; governmental organizations such as the U.S. Army Corps and the Federal Highway Administration; certification labs such as TRI Environmental; industry research labs such as Owen Cornings; manufacturers such as the Doha Waterproof Factory; consulting firms such as Advanced Earth Sciences; and academic institutions. GSI fills the same roles the AI consortium seeks to fill. It provides accreditation via inspections and testing for labs, certifies products, and writes specifications and standards. The GSI structure is composed of five separate branches responsible for different aspects of its work. The consortium could adopt this model and have separate branches within its operations that specialize in the consortia's different objectives. GSI includes the Geosynthetic Education Institute (GEI), the Geosynthetic Research Institute (GRI), the Geosynthetic Information Institute (GII), the Geosynthetic Certification Institute (GCI), and the Geosynthetic Accreditation Institute (GAI). GSI's research arm conducts product research that is of interest to its own knowledge of shortcomings in the field, with the aim that the research informs its evaluation program. The consortium could adopt a similar endeavor—given its knowledge and authority in the evaluations landscape, it could identify and solve problems with evaluation methods for example that do not receive sufficient attention in for-profit labs.

Quality Assurance International (QAI) certifies organic products for a variety of state agricultural oversight programs, such as the USDA National Organic Program (NOP), the Mexico Organics Product Law Standard (LPO), the Canadian Organic Regime (COR), the Quebec Organic Reference Standards; European Union Organic Regulation, and NSF/ANSI 305 Organic Personal Care Products (QAI International 2023). QAI decides on what the organic standards are, conducts inspections of farms desiring to be certified and then has the authority to give or withhold the label. They can certify procedures, products, farms, distributors, and restaurants and their enforcement mechanism is withholding or revoking the label. The partnering states have the ability to allow an uncertified product to be distributed to their market; however, they lack the authority to place the QAI label on uncertified products. The consortium would play a similar role as QAI in the frontier AI risk evaluations space. The difference is that the consortium would additionally need to certify evaluators, since it does not plan to conduct inspections itself due to bandwidth and transparency issues. Further, it would be ideal for the consortium to extend enforcement authority beyond its own stamp. GSI and QAI both rely on commitments from partner states to enforce their decisions. While this has worked robustly for partnering institutions, AI evaluations have an additional coordination challenge: it only works if a vast number of nations choose to participate. A country allowing falsely stamped organic products into its market does not compromise the ability of a participating state to prohibit them. But with safety, the actions of non-partner states can impact the ability of partner states to stay safe. To accomplish broad international partnerships will likely require the cooperation of other international AI organizations that should have membership in the proposed consortium.

## 4.5 Alternatives to a Consortium

This paper has proposed an international consortium for AI risk evaluations that builds on the regulatory markets approach proposed by Hadfield and Clark (2023) to incorporate an additional regulatory intermediary (Abbott et al. 2017) in the form of a consortium to coordinate evaluations efforts, set standards, and provide a competitive certification process on an international scale that individual nations would not be able to effectively incentivize. While this approach seems both practical and tractable, a consortium is not necessarily the only solution. The most important lessons of this paper do not involve the need for a consortium, but rather, they concern the need for coordination in AI risk evaluations—a sentiment echoed by others for broader AI governance efforts (Gutierrez 2023; Askell et al. 2019)—and the need for standards setting and risk response mechanisms to be integrated with risk evaluations to appropriately manage risks from dangerous emergent capabilities.

There are a number of institutional options for the governance of frontier AI that could be viable alternatives to the proposed consortium (Ho et al., 2023; Anderljung et al, 2023). Multilateral AI safety organizations could fill a similar role, promoting standards setting, international norms, and responsible AI regulations. A lightweight alternative with a number of successful precedents could be an international commission on frontier AI evaluations. In the same vein as the Intergovernmental Panel on Climate Change (IPCC), an international commission on frontier AI evaluations would act as a central body to inculcate expert consensus on AI risk and provide assessments to policymakers on emergent risks from frontier systems, appropriate evaluation metrics and collaborate with industry and academia on over-the-horizon risks. The commission could act as a hub for AI epistemic communities to centralize knowledge and communicate with one voice to the international community. However, preventing extreme risks from next-generation systems will likely require more than expert advisement and agreement on AI risk evaluation best practices.

# 5 Challenges and Limitations

In this paper we outline the need for an international consortium for frontier AI risk evaluation, and in the previous section we draw lessons from existing organizations' functions and governance structures to explore how such an organization might operate and be governed. However, this project is at an early stage and further research and reflection will be needed. Below we discuss key limitations and challenges not discussed in previous sections.

The proposed consortium is seeking to solve a coordination problem whose solution is necessary to prevent a broad range of social and national security risks. Therefore, any solution will inherently require cooperation of all stakeholders. This may pose a significant challenge, as firms may be hesitant about third-party evaluations due to challenges they present (Anthropic 2023b). Google Deepmind goes as far as to suggest that capabilities evaluations are best suited for internal teams with minimal involvement of third-party stakeholders—government regulators or third-party evaluators (Weidinger et al. 2023). The paper further implies that unreleased proprietary systems will involve novel components that evaluators and developers will need to coalesce around—something that won't be possible for government regulators or third-party evaluators—in order to ensure the most reliable safety assurances. While such circumstances may pose novel challenges that are difficult to manage, the argument rings hollow. Considering the IAEA, there is precedent for external auditing of sensitive technologies with equally novel components that are sometimes air gapped in hardened facilities to protect from kinetic attacks. These are extreme challenges to any auditing regime, yet, states are able to accommodate IAEA inspections. No excuse should be afforded frontier AI developers that want to exclude the public from the process of ensuring public safety at a global scale.

Starting an international organization with a broad mandate may be particularly challenging without widespread stakeholder support. Therefore, without such support it could be necessary to initially narrow the scope of the consortium's mandate. While a narrowing of the initial scope carries significant risks of failure, plausibly, choosing to first specialize in one domain to demonstrate viability may be necessary to attract a critical mass of evaluators to achieve more ambitious goals. Potential options for such a narrowing of the scope might be a small proof-of-concept effort to demonstrate the effectiveness of coordination with a group of supportive evaluators and at least one supportive frontier AI lab, or, an effort to begin with the creation of standards. Beginning with the creation of standards may make the most sense if only there is reason to focus only on a single objective[64]; however, a narrowing of the scope would not necessarily require focusing on a single objective, and efforts toward each of these options may be tractable and less prone to the possible failure to ever move beyond a single narrow objective.

Given the many different efforts being proposed for international AI governance, including adaptation of existing international institutions (Maas and Villalobos 2023), it is important that the proposed consortium strives to complement or supplement the new or existing international institutions that are eventually tasked with significant regulatory roles involving societal-scale AI risks. While different regulatory frameworks are being actively discussed, and it is unclear as to which one(s) will emerge favorably, we feel that we have made a strong case for why the proposed consortium will provide additional value beyond other salient proposals. However, if other frameworks emerge with consensus, the consortium may need to harmonize with those.

One particular concern is the training of open source systems (Guzman 2023), and the consortium would need to be able to facilitate the capacity to evaluate any open source projects training advanced AI systems that would constitute either fringe or frontier systems. However, it would not be the purview of the proposed consortium to ensure that all actors training systems that meet the criteria of either frontier or fringe systems are working through the proper channels to ensure verifiably safe system development. Open source projects are particularly risky because even with regulators' cooperation, there are few mechanisms in place to monitor the development of models that are outside regulators' traditional horizon. Moreover, rogue states may also try training models outside the scrutiny of international bodies. Thus, complementary regulatory functions are necessary for minimizing the proliferation of unsafe advanced AI systems; proposals for compute governance (Shavit 2023) have the potential to mitigate risks along these lines, but much additional work is needed to build the infrastructure and cooperation necessary to see them realized.

Another limitation involves the challenges associated with scaling an international organization. While a national organization would be less likely to foster collaboration and inclusion at the scale of an international organization, there are benefits, such as being easier to set up and scale, and being more flexible and having greater enforcement power. These are some of the most significant challenges of an international consortium, and the possibility of a national consortium is something that could be considered. Additionally, a national consortium or multiple national consortia could make it easier to coordinate at an international scale if the consortium or consortia were willing to coordinate efforts.

---

[64] For a coordinating body to be successful, there must be trustworthy agreed-upon standards for all stakeholders.

There are various competing challenges between the regulatory markets approach and the alternative approach(es). A discussion of these challenges is beyond the scope of this work, but is something that should be explored expeditiously while soliciting feedback from stakeholders and taking other steps described in the plan for action in the section below.

## 5.1 Potential Failure Modes

Moving forward to realize an organization resembling the proposed consortium, it will be prudent to remain cognizant of potential failure modes. Some examples include the proposed consortium being launched with the best intentions but ultimately gravitating away from the original intent.

One way that this could happen is if the consortium devolves into a lobbying group or advocacy group. If too much power within the consortium is ceded to independent AI risk evaluators or AI developers. This could lead to the consortium advocating or leveraging its connections with policymakers on behalf of AI developers or evaluators while doing little to address the concerns about the status quo set forth here. In order to mitigate this, it is imperative that the governing body of the consortium is composed of a diverse group of policymakers, risk governance experts, AI experts, academics, and domain experts in the capabilities being evaluated. Attracting and retaining this group with its diverse expertise will be a challenge due to competition with AI developers, third-party evaluators, and other groups wishing to make use of this expertise, particularly since many of these experts are able to command large salaries that the consortium may not be able to match. Additionally, their members could be assigned different roles, e.g., as an observer or participant.

A second mode of failure for the proposed consortium is that it becomes a standards-setting organization that lacks the levers necessary for enforcement. This could occur if coordination with regulators is insufficient or ineffective, or, if the consortium becomes, or is perceived to be, primarily focused on the West or global north. Genuine engagement with stakeholders from outside of these areas will be imperative both for the consortium to understand relevant cultural considerations when engaging with local policymakers and experts, as well as for giving groups outside of these areas a greater voice in shaping AI evaluation. The third way in which the consortium could become unable to enforce its decisions is if the expertise or focus of the consortium lags too far behind the frontier of industry. If this were the case, proposals by the consortium may be simply irrelevant to the most prescient societal-scale risks. In order to mitigate this, efforts are needed to ensure that the consortium remains up to date with cutting edge systems and advances in evaluation science, and AI governance proposals.

Literature in related fields details other ways a consortium could fail. Technological patent consortia have been known to fail when there is insufficient incentive for firms to participate (Liz-Gutierrez et al. 2016). This could be a particular problem in evaluations, where key labs only participate with the goal of shaping the evaluation space toward their own interest. There is also a danger of 'free-rider' firms who participate with the goal of gaining knowledge from others rather than contributing. Blockchain consortia have been known to fail due to not adapting operations strategies as the consortium grows, and due to attrition of the members who do not have the desire to stay long-term (Massey et al. 2020). Research consortia who grow overly dependent on the outsize contribution of one member can fail if that member leaves, or can fail if its members cannot reach consensus on goals and priorities (Kherrazi 2023). This could be similarly problematic for an evaluations consortium if there was not enough diversity such that multiple evaluators were specializing in evaluating each of the various societal-scale AI risks of concern. On a higher level, international organizations in general have historically failed for two primary reasons (Eilstrup-Sangiovanni 2020). One is exogenous power shifts or shocks such as economic crises and war. These increase the costs of cooperation, or may create new realities that are no longer amenable to old agreements. The other is that the international organization simply failed to grow after inception. Their membership remained small and never developed a strong, centralized governance. Even when international organizations don't fail, they can become what Gray (2018) terms 'zombies', that is, organizations that still exist but no longer make real progress towards their mandates. The two main contributors to 'zombie' status are staff attrition, when the most talented or valuable staff leave and cannot be replaced, and politicization, when state members wield too much power and the organization becomes their political vehicle rather than an independent body acting for the good of their mission.

Finally, it is possible that a consortium could exacerbate current problems if it were to make coordination more challenging by adding an increased layer of communication. Additionally, it could be that a consortium, and the lack of centralization of evaluations, could have a negative effect—some have proposed that in-house evaluations by governments are preferable to the third-party approach (Whittlestone and Clark 2021).

# 6 Conclusion: A Plan for Action

We have described challenges to evaluating advanced AI systems for susceptibility to societal-scale risks. As a solution we have proposed the creation of an international consortium for advanced AI risk evaluations. We have discussed the objectives of the proposed consortium, issues related to its operations, matters related to its governance, and how it relates to existing and proposed international institutions, as well as a discussion of challenges and limitations. In conclusion, we provide a plan of action for moving forward with the proposal, involving the following next steps:

1. Collect feedback from AI researchers, think tanks, and potential stakeholders
2. Continue to conduct research to better understand similar existing organizations
3. Conduct a workshop with a diverse group of experts and future stakeholders to explore how the consortium could be structured
4. Assemble a new team to incorporate elicited feedback on the proposal to aid in preparing a detailed plan of action for quickly establishing and scaling a consortium
5. Solicit funding for the detailed plan of action
6. Form the consortium
    - Select an advisory board, board of directors, and executive team
    - Have evaluators join the consortium and begin operations

# Contributions

Ross Gruetzemacher and Siméon Campos provided initial inspiration. Ross Gruetzemacher, Alan Chan, Jose Hernández-Orallo, and Christy Manning helped guide the shape and direction of the project. Ross Gruetzemacher, Christy Manning, Štěpán Los, Kevin Frazier, Alan Chan, and Kyle Kilian contributed to research. Ross Gruetzemacher, Christy Manning, Alan Chan, Kevin Frazier, Kyle Kilian, and Štěpán Los were responsible for writing significant sections. These and all other authors participated in meetings, contributed substantively to writing, or provided substantive suggestions and comments.

# Acknowledgements

We thank David Manheim, Oliver Guest, Philip Tomei, Sammy Martin, Lennart Heim, Matthijs Maas, and Sean Wissing for comments and discussions at different stages of the project. Direct support was provided by the Transformative Futures Institute.

# References

Acemoglu, D., and Johnson, S. 2023. Power and progress: our thousand-year struggle over technology and prosperity. PublicAffairs, New York, first edition, 2023. ISBN 9781541702530.

Adiwardana, D., Luong, M.T., So, D.R., Hall, J., Fiedel, N., Thoppilan, R., Yang, Z., Kulshreshtha, A., Nemade, G., Lu, Y. and Le, Q.V., 2020. Towards a human-like open-domain chatbot. arXiv preprint arXiv:2001.09977.

Airplane Owners and Pilots Association, 2019. State of General Aviation Report. https://download.aopa.org/hr/Report_on_General_Aviation_Trends.pdf

Alaga, J., Schuett, J. 2023. Coordinated pausing: An evaluation-based coordination scheme for frontier AI developers. Center for the Governance of AI.

Alignment Research Center, Evaluations Team (ARC Evals). 2023. Responsible Scaling Policies. Alignment Research Center, blog. https://evals.alignment.org/blog/2023-09-26-rsp/

Anderljung, M., Barnhart, J., Leung, J., Korinek, A., O'Keefe, C., Whittlestone, J., Avin, S., Brundage, M., Bullock, J., Cass-Beggs, D. and Chang, B., 2023. Frontier AI Regulation: Managing Emerging Risks to Public Safety. arXiv preprint arXiv:2307.03718.

Anthropic. 2023a. Anthropic's Responsible Scaling Policy. https://www.anthropic.com/index/anthropics-responsible-scaling-policy

Anthropic. 2023b. Challenges in Evaluating AI Systems. https://www.anthropic.com/index/evaluating-ai-systems

Anthropic. 2023c. Frontier Threats Red Teaming for AI Safety. https://www.anthropic.com/index/frontier-threats-red-teaming-for-ai-safety

Anthropic. 2023d. Model Card Evaluations for Claude Models. https://efficient-manatee.files.svdcdn.com/production/images/Model-Card-Claude-2.pdf

Armstrong, S., Bostrom, N. and Shulman, C., 2016. Racing to the precipice: a model of artificial intelligence development. AI & society, 31, pp.201-206.

Askell, A., Brundage, M. and Hadfield, G., 2019. The role of cooperation in responsible AI development. arXiv preprint arXiv:1907.04534.

Auld, G., Bernstein, S., Cashore, B. and Levin, K., 2021. Managing pandemics as super wicked problems: lessons from, and for, COVID-19 and the climate crisis. Policy sciences, 54, pp.707-728.

Avin. S. 2023. Frontier AI Regulation Blueprint. Blog. Centre for the Study of Existential Risk, University of Cambridge. https://www.cser.ac.uk/news/frontier-ai-regulation-blueprint/

Bengio, Y., 2023. AI and Catastrophic Risk. Journal of Democracy, 34(4), pp.111-121.

Boiko, D.A., MacKnight, R. and Gomes, G., 2023. Emergent autonomous scientific research capabilities of large language models. arXiv preprint arXiv:2304.05332.

Bommasani, R., Hudson, D.A., Adeli, E., Altman, R., Arora, S., von Arx, S., Bernstein, M.S., Bohg, J., Bosselut, A., Brunskill, E. and Brynjolfsson, E., 2021. On the opportunities and risks of foundation models. arXiv preprint arXiv:2108.07258.

Bommasani, R., Klyman, K., Longpre, S., Kapoor, S., Maslej, N., Xiong, B., Zhang, D., and Liang, P. 2023. The Foundation Model Transparency Index. arXiv.2310.12941v1.

Bresnahan, T.F. and Trajtenberg, M., 1995. General purpose technologies 'Engines of growth'?. Journal of econometrics, 65(1), pp.83-108.

Brundage, M., Avin, S., et al. 2020. Toward Trustworthy AI Development: Mechanisms for Supporting Verifiable Claims. arXiv:2004.07213.

Bubeck, S., Chandrasekaran, V., Eldan, R., Gehrke, J., Horvitz, E., Kamar, E., Lee, P., Lee, Y.T., Li, Y., Lundberg, S. and Nori, H., 2023. Sparks of artificial general intelligence: Early experiments with gpt-4. arXiv preprint arXiv:2303.12712.

Burnell, R., Burden, J., Rutar, D., Voudouris, K., Cheke, L., & Hernández-Orallo, J., 2022. Not a Number: Identifying Instance Features for Capability-Oriented Evaluation. In IJCAI (pp. 2827-2835).

Burnell, R., Schellaert, W., Burden, J., Ullman, T. D., Martinez-Plumed, F., Tenenbaum, J. B., ... and Hernandez-Orallo, J., 2023. Rethink reporting of evaluation results in AI. *Science*, *380*(6641), 136-138.

Chan, A., Salganik, R., Markelius, A., Pang, C., Rajkumar, N., Krasheninnikov, D., Langosco, L., He, Z., Duan, Y., Carroll, M. and Lin, M., 2023, June. Harms from Increasingly Agentic Algorithmic Systems. In Proceedings of the 2023 ACM Conference on Fairness, Accountability, and Transparency (pp. 651-666).

Chin, C. 2023. Navigating the Risks of Artificial Intelligence on the Digital News Landscape. Blog. CSIS. https://www.csis.org/analysis/navigating-risks-artificial-intelligence-digital-news-landscape

Christian, B., 2020. The alignment problem: Machine learning and human values. WW Norton & Company.

Cihon, P., Kleinaltenkamp, M. J., Schuett, J., and Baum, S. D. 2021. AI certification: Advancing ethical practice by reducing information asymmetries. In *IEEE Transactions on Technology and Society*. LPP Working Paper No. 10, 2021.

Clark, Jack, and Gillian K. Hadfield. "Regulatory markets for AI safety." arXiv preprint arXiv:2001.00078 (2019).

Costanza-Chock, S., Raji, I. D. and Buolamwini, J. 2022. Who Audits the Auditors? Recommendations from a field scan of the algorithmic auditing ecosystem. In Proceedings of the 2022 ACM Conference on Fairness, Accountability, and Transparency (pp. 1571–1583).

Critch, A. and Russell, S., 2023. TASRA: A Taxonomy and Analysis of Societal-Scale Risks from AI. arXiv preprint arXiv:2306.06924.

Dafoe, A., 2018. AI governance: a research agenda. Governance of AI Program, Future of Humanity Institute, University of Oxford: Oxford, UK, 1442, p.1443.

Davidson, T., 2023. The Danger of Runaway AI. Journal of Democracy, 34(4), pp.132-140.

Deng, J., Dong, W., Socher, R., Li, L.J., Li, K. and Fei-Fei, L., 2009, June. Imagenet: A large-scale hierarchical image database. In 2009 IEEE conference on computer vision and pattern recognition (pp. 248-255). IEEE.

Eilstrup-Sangiovanni, M. 2020. What kills international organizations? When and why international organisations terminate. European Journal of International Relations, 27(1).

Eloundou, T., Manning, S., Mishkin, P. and Rock, D., 2023. GPTs are GPTs: An early look at the labor market impact potential of large language models. arXiv preprint arXiv:2303.10130.

Engler, A., 2023. A comprehensive and distributed approach to AI regulation. Brookings Institute. August 31, 2023. https://www.brookings.edu/articles/a-comprehensive-and-distributed-approach-to-ai-regulation/

European Commission, 2023. Sectorial AI Testing and Experimentation Facilities under the Digital Europe Programme. https://digital-strategy.ec.europa.eu/en/activities/testing-and-experimentation-facilities

European Union. (2021). Regulation on laying down harmonised rules on artificial intelligence (Artificial Intelligence Act) and amending certain Union legislative acts. Official Journal of the European Union, L 344/1.

Farmer, J. D., Hepburn, C., Ives, M. C., Hale, T., Wetzer, T., Mealy, P., ... & Way, R. (2019). Sensitive intervention points in the post-carbon transition. Science, 364(6436), 132-134.

Fist, T., Depp, M., Withers, C., 2023. Response to OSTP "National Priorities for Artificial Intelligence Request for Information". Center for a New American Security. July 20, 2023. https://www.cnas.org/publications/commentary/ostp-national-priorities-for-artificial-intelligence

Future of Life Institute, 2023. Pause Giant AI Experiments: An Open Letter. https://futureoflife.org/open-letter/pause-giant-ai-experiments/

Ganguli, D., Hernandez, D., Lovitt, L., Askell, A., Bai, Y., Chen, A., Conerly, T., Dassarma, N., Drain, D., Elhage, N., El Showk, S., et al. 2022, June. Predictability and surprise in large generative models. In Proceedings of the 2022 ACM Conference on Fairness, Accountability, and Transparency (pp. 1747-1764).

Goodhart, C.A.E., 1984. Problems of monetary management: the UK experience (pp. 91-121). Macmillan Education UK.

GOV.UK, 2022. Auditing algorithms: the existing landscape, role of regulators and future outlook. https://www.gov.uk/government/publications/findings-from-the-drcf-algorithmic-processing-workstream-spring-2022/auditing-algorithms-the-existing-landscape-role-of-regulators-and-future-outlook

Gray, J. 2018. Life, death, or zombie? The vitality of international organizations. International Studies Quarterly, 62(1), 1-13.

Gruetzemacher, R. 2018. Rethinking AI Strategy and Policy as Entangled Super Wicked Problems. http://www.rossgritz.com/wp-content/uploads/2018/11/aies_gruetzemacher_revisions.pdf

Gruetzemacher, R. and Whittlestone, J., 2022. The transformative potential of artificial intelligence. Futures, 135, p.102884.

Gruetzemacher, R. and Paradice, D., 2022. Deep transfer learning & beyond: Transformer language models in information systems research. ACM Computing Surveys (CSUR), 54(10s), pp.1-35.

Gutierrez, C.I., 2023. Multilateral Coordination for the Proactive Governance of Artificial Intelligence Systems. Available at SSRN.

Gutierrez, C. I., Aguirre, A., Uuk, R., Boine, C. C., & Franklin, M. (2023). A proposal for a definition of general purpose artificial intelligence systems. Digital Society, 2(3), 36.

Guzman, H. 2023. Planned Protest Throws the Spotlight on Risks, Rewards of Open-Source AI. Law.com. https://www.law.com/corpcounsel/2023/09/28/planned-protest-throws-the-spotlight-on-risks-rewards-of-open-source-ai/

Habuka, H., 2023. The Path to Trustworthy AI: G7 Outcomes and Implications for Global AI Governance. Center for Strategic and International Studies. June 6, 2023. https://www.csis.org/analysis/path-trustworthy-ai-g7-outcomes-and-implications-global-ai-governance

Hadfield, G.K. and Clark, J., 2023. Regulatory Markets: The Future of AI Governance. arXiv preprint arXiv:2304.04914.

Hao, K. 2020. We read the paper that forced Timnit Gebru out of Google. Here's what it says. MIT Tech Review. https://www.technologyreview.com/2020/12/04/1013294/google-ai-ethics-research-paper-forced-out-timnit-gebru/

Harris, S., Suleyman, M. 2023. Can We Contain Artificial Intelligence. Podcast 322. August, 2023. https://www.samharris.org/podcasts/making-sense-episodes/332-can-we-contain-artificial-intelligence

Hendrycks, D., Burns, C., Basart, S., Zou, A., Mazeika, M., Song, D. and Steinhardt, J., 2021. Measuring massive multitask language understanding. International Conference on Learning Representations.

Hendrycks, Dan, Mantas Mazeika, and Thomas Woodside. 2023. "An Overview of Catastrophic AI Risks." arXiv preprint arXiv:2306.12001

Hernández-Orallo, J., 2017a. The measure of all minds: evaluating natural and artificial intelligence. Cambridge University Press.

Hernández-Orallo, J. 2017b. Evaluation in artificial intelligence: from task-oriented to ability-oriented measurement. Artificial Intelligence Review, 48, 397-447.

Ho, L., Barnhart, J., Trager, R., Bengio, Y., Brundage, M., Carnegie, A., Chowdhury, R., Dafoe, A., Hadfield, G., Levi, M. and Snidal, D., 2023. International Institutions for Advanced AI. arXiv preprint arXiv:2307.04699

Hobbhahn, M., and Besiroglu, T. 2022. Trends in GPU price-performance. Blog. https://epochai.org/blog/trends-in-gpu-price-performance

Huang, Y., Gupta, S., Xia, M., Li, K., and Chen, D. 2023. Catastrophic jailbreak of open-source LLMs via exploiting generation. https://arxiv.org/pdf/2310.06987.pdf

Hubinger, E., van Merwijk, C., Mikulik, V., Skalse, J. and Garrabrant, S., 2019. Risks from learned optimization in advanced machine learning systems. arXiv preprint arXiv:1906.01820.

Institute of Internal Auditors (IIA). 2018. Global Perspectives and Insights: The IIA's AI Auditing Framework. https://www.theiia.org/globalassets/documents/content/articles/gpi/2017/december/gpi-artificial-intelligence-part-ii.pdf

Institute of Electrical and Electronics Engineers: Industry Standards and Technology Organization. March 8th, 2023. 5 reasons why you should start a consortium. https://ieee-isto.org/?s=5+reasons+why

ISACA, 2018. Auditing Artificial Intelligence. https://ec.europa.eu/futurium/en/system/files/ged/auditing-artificial-intelligence.pdf

Jaiswal, S. D., Verma A., and Mukherjee, A. 2023. Auditing gender analyzers on text data. https://arxiv.org/pdf/2310.06061.pdf

Jelinek, T., Wallach, W. and Kerimi, D., 2021. Policy brief: the creation of a G20 coordinating committee for the governance of artificial intelligence. AI and Ethics, 1(2), pp.141-150.

Kherrazi, S. 2023. Exploring taxonomies and governance challenges of sponsored R&D consortia: Evidence from the EU framework program. Journal of Innovation Economics and Management, 41(2), 217-249.

Kinniment, M., Sato, L.J.K., Du, H., Goodrich, B., Hasin, M., Chan, L., Miles, L.H., Lin, T.R., Wijk, H., Burget, J., Ho, A., Barnes, E. and Christiano, P. 2023. Evaluating Language-Model Agents on Realistic Autonomous Tasks. Alignment Research Center, Evaluations Team (ARC Evals).

Lazar, S. and Nelson, A., 2023. AI safety on whose terms?. Science, 381(6654), pp.138-138.

Levesque, H., Davis, E. and Morgenstern, L., 2012, May. The winograd schema challenge. In Thirteenth international conference on the principles of knowledge representation and reasoning.

Levin, K., Cashore, B., Bernstein, S. and Auld, G., 2012. Overcoming the tragedy of super wicked problems: constraining our future selves to ameliorate global climate change. Policy sciences, 45(2), pp.123-152.

Liang, P., Bommasani, R., Lee, T., Tsipras, D., Soylu, D., Yasunaga, M., Zhang, Y., Narayanan, D., Wu, Y., Kumar, A. and Newman, B., 2023. Holistic evaluation of language models. Transactions on Machine Learning Research.

Lieke, J., and Sutskever, I. 2023. Introducing Superalignment. OpenAI, blog. https://openai.com/blog/introducing-superalignment

Lipsey, R.G., Carlaw, K.I. and Bekar, C.T., 2005. Economic transformations: general purpose technologies and long-term economic growth. Oup Oxford.

Lin, S., Hilton, J., and Evans, O. 2022. Truthful QA: Measuring how models mimic human falsehoods. In Proceedings of the 60th Annual Meeting of the Association for Computational Linguistics (Volume 1: Long Papers), pages 3214–3252, Dublin, Ireland. Association for Computational Linguistics.

Lis-Gutiérrez, J. P., Lis-Gutiérrez, M., Viloria, A., Gaitán-Angulo, M., and Balaguera, M. I. 2016. Main failure factors in technology consortia. International Journal of Control Theory and Applications, 9(44), 395-399.

McCarthy, J., Minsky, M., Rochester, N. and Shannon, C., 1956. Dartmouth Summer Research Conference on Artificial Intelligence. Dartmouth College.

Maas, M. 2023. Elements of Advanced AI Governance: A Literature Review of Key Concepts and Terms. _AI Foundations Report 2. Legal Priorities Project.

Maas, M.M. and Villalobos, J.J., 2023. International AI Institutions: A Literature Review of Models, Examples, and Proposals. AI Foundations Report, 1.

Manheim, D. and Garrabrant, S., 2018. Categorizing variants of Goodhart's Law. arXiv preprint arXiv:1803.04585.

Manheim, D., 2023. Building less-flawed metrics: Understanding and creating better measurement and incentive systems. Patterns, 4(10).

Massey, R., Prokop, M., Henry, W., Taylor, P., and Simpson, L. 2020. Governance and structuring considerations in blockchain consortia. Deloitte, report. https://www2.deloitte.com/content/dam/Deloitte/us/Documents/strategy/us-deloitte-governance-structuring-considerations-blockchain-consortia.pdf

Matteucci, K., Avin, S., Barez, F. and hÉigeartaigh, S.Ó., 2023. AI Systems of Concern. arXiv preprint arXiv:2310.05876.

Meadows, Donella. H. 2008. Thinking in Systems: A Primer. Chelsea Green Publishing.

Mökander, J., Schuett, J., Kirk, H.R. and Floridi, L., 2023. Auditing large language models: a three-layered approach. AI and Ethics, pp.1-31.

Montalbano, E. 'DarkBERT' GPT-Based Malware Trains Up on the Entire Dark Web. DarkReading. https://www.darkreading.com/application-security/gpt-based-malware-trains-dark-web

Mouton, C., Lucas, C., and Guest, E. 2023. The Operational Risks of AI in Large-scale Biological Attacks: A Red-team Approach. RAND Corporation. https://www.rand.org/pubs/research_reports/RRA2977-1.html

National Academies of Sciences, Engineering, and Medicine. 2023. Test and Evaluation Challenges in Artificial Intelligence-Enabled Systems for the Department of the Air Force. Washington, DC: The National Academies Press. https://doi.org/10.17226/27092

National Artificial Intelligence Research Resource (NAIRR) Task Force, 2023. Strengthening and Democratizing the U.S. Artificial Intelligence Innovation Ecosystem: An Implementation Plan for a National Artificial Intelligence Research Resource. NAIRR Task Force. https://www.ai.gov/wp-content/uploads/2023/01/NAIRR-TF-Final-Report-2023.pdf

National Institute of Standards and Technology. 2023. Artificial Intelligence Risk Management Framework (AI RMF 1.0). United States Department of Commerce. https://nvlpubs.nist.gov/nistpubs/ai/NIST.AI.100-1.pdf

OECD (2022). OECD Framework for the Classification of AI Systems. OECD Digital Economy Papers. No. 323, OECD Publishing, Paris. https://doi.org/10.1787/cb6d9eca-en

OpenAI. 2023a. GPT-4 technical report. arXiv, pp.2303-08774.

OpenAI. 2023b. OpenAI Red Teaming Network. https://openai.com/blog/red-teaming-network

OpenAI. 2023c. Frontier Model Forum. A joint announcement from Anthropic, OpenAI, Microsoft, and Google. https://openai.com/blog/frontier-model-forum

Oremus, W. September 16th, 2021. Facebook keeps researching its own harms - and burying the findings. The Washington Post. https://www.washingtonpost.com/technology/2021/09/16/facebook-files-internal-research-harms/

Paglia, E. and Parker, C., 2021. The intergovernmental panel on climate change: guardian of climate science. Guardians of Public Value: How Public Organisations Become and Remain Institutions, pp.295-321.

Patel, D., Wong, G. 2023. GPT-4 Architecture, Infrastructure, Training Dataset, Costs, Vision, MoE. Blog. Semi Analysis, https://www.semianalysis.com/p/gpt-4-architecture-infrastructure

Perry, B. and Uuk, R., 2019. AI governance and the policymaking process: key considerations for reducing AI risk. Big data and cognitive computing, 3(2), p.26.

Pichai, S. 2023. Google I/O 2023: Making AI more helpful for everyone. https://blog.google/technology/ai/google-io-2023-keynote-sundar-pichai/#palm-2-gemini

Raji, I.D., Bender, E.M., Paullada, A., Denton, E. and Hanna, A., 2021. AI and the everything in the whole wide world benchmark. arXiv preprint arXiv:2111.15366.

Raji, I. D., Xu P., Honigsberg C., Ho D. E. 2022a. Outsider Oversight: Designing a Third Party Audit Ecosystem for AI Governance. arXiv: 2206.04737 [cs.CY].

Raji, I.D., Kumar, I.E., Horowitz, A. and Selbst, A., 2022b, June. The fallacy of AI functionality. In Proceedings of the 2022 ACM Conference on Fairness, Accountability, and Transparency (pp. 959-972).

Rittel, H.W. and Webber, M.M., 1973. Dilemmas in a general theory of planning. Policy sciences, 4(2), pp.155-169.

Roller, S., Dinan, E., Goyal, N., Ju, D., Williamson, M., Liu, Y., Xu, J., Ott, M., Shuster, K., Smith, E.M. and Boureau, Y.L., 2021. Recipes for building an open-domain chatbot. EACL Conference 2021.

Sætra, H.S. and Danaher, J., 2022. To each technology its own ethics: The problem of ethical proliferation. Philosophy & Technology, 35(4), p.93.

ScaleAI. 2023. Testing and Evaluation Vision. ScaleIA, blog. https://scale.com/guides/test-and-evaluation-vision#vision-for-the-t&e-ecosystem

Schuett, J., Dreksler, N., Anderljung, M., McCaffary, D., Heim, L., Bluemke, E. and Garfinkel, B., 2023. Towards best practices in AGI safety and governance: A survey of expert opinion. arXiv preprint arXiv:2305.07153.

Searle, J.R., 1980. Minds, brains, and programs. Behavioral and brain sciences, 3(3), pp.417-424.

Shah, A. 2023. Google TPU v5e AI Chip Debuts after Controversial Origins. HPC Wire. https://www.hpcwire.com/2023/08/30/google-tpu-v5e-ai-chip-debuts-after-controversial-origins/

Shavit, Y., 2023. What does it take to catch a Chinchilla? Verifying Rules on Large-Scale Neural Network Training via Compute Monitoring. arXiv preprint arXiv:2303.11341.

Shevlane, T., 2022. Structured access: an emerging paradigm for safe AI deployment. arXiv preprint arXiv:2201.05159.

Shevlane, T., Farquhar, S., Garfinkel, B., Phuong, M., Whittlestone, J., Leung, J., Kokotajlo, D., Marchal, N., Anderljung, M., Kolt, N. and Ho, L., 2023. Model evaluation for extreme risks. arXiv preprint arXiv:2305.15324.

Singh, J.P., Amarda Shehu, Caroline Wesson, and Manpriya Dua. 2023. The 2023 Global Artificial Intelligence Infrastructures Report. With a Foreword from David Bray. AI Strategies Team and the Institute for Digital Innovation, George Mason University, and the Stimson Center, Washington DC. August 2023

Srivastava, A., Rastogi, A., Rao, A., Shoeb, A.A.M., Abid, A., Fisch, A., Brown, A.R., Santoro, A., Gupta, A., Garriga-Alonso, A. and Kluska, A., et al. 2022. Beyond the imitation game: Quantifying and extrapolating the capabilities of language models.*Transactions on Machine Learning*.

Stern, P., 2011. Design principles for global commons: Natural resources and emerging technologies. International Journal of the Commons, 5(2).

Suleyman, M., Bhaskar, M., 2023. The Coming Wave. Penguin Random House.

Sullivan, W. 1987. British to Remain in Research Group. New York Times. December 21, 1987. https://www.nytimes.com/1987/12/21/us/british-to-remain-in-research-group.html

Tabassi, E. (2023). Artificial Intelligence Risk Management Framework (AI RMF 1.0). National Institute of Standards and Technology. https://doi.org/10.6028/NIST.AI.100-1

Teney, D., Abbasnejad, E., Kafle, K., Shrestha, R., Kanan, C. and Van Den Hengel, A., 2020. On the value of out-of-distribution testing: An example of goodhart's law. Advances in neural information processing systems, 33, pp.407-417.

The White House, Office of the Press Secretary. May 4, 2023. FACT SHEET: Biden-Harris Administration Announces New Actions to Promote Responsible AI Innovation that Protects Americans' Rights and Safety https://www.whitehouse.gov/briefing-room/statements-releases/2023/05/04/fact-sheet-biden-harris-administration-announces-new-actions-to-promote-responsible-ai-innovation-that-protects-americans-rights-and-safety/

Trager, R., Harack, B., Reuel, A., Carnegie, A., Heim, L., Ho, L., Kreps, S., Lall, R., Larter, O., Ó hÉigeartaigh, S., and Staffell, S., 2023. International Governance of Civilian AI: A Jurisdictional Certification Approach. arXiv preprint arXiv:2308.15514.

Turing, A., 1950. Turing. Computing machinery and intelligence. Mind, 59(236), pp.433-60.

U.S. Senate Subcommittee on Privacy, Technology and the Law, 2023. "Oversight of A.I.: Principles for Regulation." Recording of hearing, July 25th, 2023. https://www.judiciary.senate.gov/committee-activity/hearings/oversight-of-ai-principles-for-regulation

U.S. Department of Energy. 2023. U.S. Department of Energy Announces Release of the Artificial Intelligence Risk Management Playbook. U.S. Department of Energy, Artificial Intelligence & Technology Office. 2023.

https://www.energy.gov/ai/articles/us-department-energy-announces-release-artificial-intelligence-risk-management-playbook

U.S. Department of Health & Human Services. Biosafety Levels. Website. https://www.phe.gov/s3/BioriskManagement/biosafety/Pages/Biosafety-Levels.aspx

Wallin, J., Drexel, B., Depp, M., Withers, C., 2023. CNAS Responds: Oversight of A.I.: Rules for Artificial Intelligence. Center for a New American Security. 2023. https://www.cnas.org/publications/commentary/ostp-national-priorities-for-artificial-intelligence

Wang, A., Singh, A., Michael, J., Hill, F., Levy, O., and Bowman, S.R. 2018. GLUE: A multi-task benchmark and analysis platform for natural language understanding. In Proceedings of EMNLP Workshop on BlackBox NLP, ACL, 353–355.

Wang, A., Pruksachatkun, Y., Nangia, N., Singh, A., Michael, J., Hill, F., Levy, O., and Bowman, S.R. 2019. SuperGLUE: A stickier benchmark for general-purpose language understanding systems. In Advances in Neural Information Processing Systems 32 (NeurIPS 2019).

Wang, X., Jiang, L., Hernandez-Orallo, J., Sun, L., Stillwell, D., Luo, F. and Xie, X., 2023. Evaluating General-Purpose AI with Psychometrics. arXiv preprint arXiv:2310.16379.

Wei, J., Tay, Y., Bommasani, R., Raffel, C., Zoph, B., Borgeaud, S., Yogatama, D., Bosma, M., Zhou, D., Metzler, D. and Chi, E.H., 2022. Emergent abilities of large language models. arXiv preprint arXiv:2206.07682.

Weidinger, L., Uesato, J., Rauh, M., Griffin, C., Huang, P.S., Mellor, J., Glaese, A., Cheng, M., Balle, B., Kasirzadeh, A. and Biles, C., 2022, June. Taxonomy of risks posed by language models. In Proceedings of the 2022 ACM Conference on Fairness, Accountability, and Transparency (pp. 214-229).

Weidinger, L., Rauh, M., Marchal, N., Manzini, A., Hendricks, L.A., Mateos-Garcia, J., et al. 2023. Sociotechnical Safety Evaluation of Generative AI Systems. Google DeepMind. arXiv:2310.11986v1.

Weizenbaum, J., 1966. ELIZA—a computer program for the study of natural language communication between man and machine. Communications of the ACM, 9(1), pp.36-45.

Whittlestone, J. and Clark, J. 2021. 31 August 2021. Why and how governments should monitor AI development. The Center for the Study of Existential Risk, Cambridge University. https://www.cser.ac.uk/resources/why-and-how-governments-should-monitor-ai-development/.

Zavlovinka, S., Ziolkowski, R., Bauer, I., and Schwabe, G. 2020. Management, governance, and value creation in a blockchain consortium. MIS Quarterly Executive, 19(1),1-17.

Zwetsloot, R. and Dafoe, A., 2019. Thinking about risks from AI: Accidents, misuse and structure. Lawfare. February, 11, p.2019.